\documentclass[a4paper,twoside]{article}
\newcommand{\affil}[1]{$^{\rm #1}$}
\usepackage[authoryear,square]{natbib}
\bibpunct{(}{)}{;}{a}{}{,}
\usepackage{psfig}
\usepackage{dcolumn}
\date{} 
\newcolumntype{d}[1]{D{.}{.}{#1}}
\def\lapp{\ifmmode\stackrel{<}{_{\sim}}\else$\stackrel{<}{_{\sim}}$\fi}
\def\gapp{\ifmmode\stackrel{>}{_{\sim}}\else$\stackrel{>}{_{\sim}}$\fi}
\def\degr{\ifmmode^{\circ}\else$^{\circ}$\fi}

\title{\large\bf\flushleft The Parkes Pulsar Timing Array Project}
\author{\parbox{\textwidth}{\flushleft
\vspace{-0.5cm}
{\it R. N. Manchester\affil{A,P}, G. Hobbs\affil{A},
  M. Bailes\affil{B}, W. A. Coles\affil{C}, 
  W. van Straten\affil{B}, M. J. Keith\affil{A}, 
  R. M. Shannon\affil{A}, N. D. R. Bhat\affil{B,D},
  A. Brown\affil{A}, S. G. Burke-Spolaor\affil{E,A},
  D. J. Champion\affil{F,A}, A. Chaudhary\affil{A},
  R. T. Edwards\affil{G}, G. Hampson\affil{A},
  A. W. Hotan\affil{A,B}, A. Jameson\affil{B}, F. A. Jenet\affil{H},
  M. J. Kesteven\affil{A}, J. Khoo\affil{A},
  J. Kocz\affil{B,I}, K. Maciesiak\affil{J,A}, S. Oslowski\affil{B,A}, 
  V. Ravi\affil{K,A},
  J. R. Reynolds\affil{A}, J. M. Sarkissian\affil{A},
  J. P. W. Verbiest\affil{F,B},
  Z. L. Wen\affil{L}, W. E. Wilson\affil{A}, D. Yardley\affil{M,A},
  W. M. Yan\affil{N}, X. P. You\affil{O}
} \\
\vspace{0.4cm}
{\small \affil{A}\,CSIRO Astronomy and Space Science, PO Box
  76, Epping NSW 1710, Australia} \\
{\small \affil{B}\,Centre for Astrophysics and Supercomputing,
  Swinburne
  University of Technology, PO Box 218, Hawthorn Vic 3122, Australia} \\
{\small \affil{C}\,Electrical \& Computer Engineering, University of
  California at San Diego, La Jolla, CA 92093, USA} \\
{\small \affil{D}\,International Centre for Radio Astronomy Research, Curtin University, 
  Bentley, WA 6102, Australia} \\
{\small \affil{E}\,Jet Propulsion Laboratory, California Institute of Technology, 4800 Oak Grove Dr, Pasadena CA 91109-8099, USA}\\
{\small \affil{F}\,Max-Planck-Institut f\"ur Radioastronomie, Auf dem
  H\"ugel 69,
  53121 Bonn, Germany} \\
{\small \affil{G}\,10 James Street, Whittlesea Vic. 3757, Australia } \\
{\small \affil{H}\,Center for Advanced Radio Astronomy, University
  of Texas at Brownsville, 80 Fort Brown, Brownsville TX 78520, USA} \\
{\small \affil{I}\,Harvard-Smithsonian Center for Astrophysics, 60
  Garden St., Cambridge MA 02138, USA} \\
{\small \affil{J}\,Kepler Institute of Astronomy, University of
  Zielona G\'ora,
  Lubuska 2, 65-265 Zielona G\'ora, Poland} \\
{\small \affil{K}\,School of Physics, University of Melbourne, Vic 3010, Australia} \\
{\small \affil{L}\,National Astronomical Observatories, CAS, Jia-20
  DaTun Road, Beijing 100012, China} \\
{\small \affil{M}\,School of Physics, University of Sydney, NSW 2006,
  Australia (Present address: 5/504 New Canterbury Road, Dulwich Hill, NSW 2203, Australia)} \\
{\small \affil{N}\,Xinjiang Astronomical Observatory, CAS, 150 Science
  1-Street, Urumqi, Xinjiang 830011, China} \\
{\small \affil{O}\,School of Physical Science \& Technology, Southwest
  University, 2 Tiansheng Road, Chongqing 400715, China} \\
{\small \affil{P}\,Email: dick.manchester@csiro.au}}}
\begin{document}
\twocolumn[
\begin{changemargin}{.8cm}{.5cm}
\begin{minipage}{.9\textwidth}
\vspace{-1cm}
\maketitle
%
%
\small{\bf Abstract:} A ``pulsar timing array'' (PTA), in which
observations of a large sample of pulsars spread across the celestial
sphere are combined, allows investigation of ``global'' phenomena such
as a background of gravitational waves or instabilities in atomic
timescales that produce correlated timing residuals in the pulsars of
the array. The Parkes Pulsar Timing Array (PPTA) is an implementation
of the PTA concept based on observations with the Parkes 64-m radio
telescope. A sample of 20 millisecond pulsars is being observed at
three radio-frequency bands, 50cm ($\sim 700$~MHz), 20cm ($\sim
1400$~MHz) and 10cm ($\sim 3100$~MHz), with observations at intervals
of 2 -- 3 weeks. Regular observations commenced in early 2005. This
paper describes the systems used for the PPTA observations and data
processing, including calibration and timing analysis. The strategy
behind the choice of pulsars, observing parameters and analysis
methods is discussed. Results are presented for PPTA data in the three
bands taken between 2005 March and 2011 March. For ten of the 20
pulsars, rms timing residuals are less than $1 \mu$s for the best band
after fitting for pulse frequency and its first time
derivative. Significant ``red'' timing noise is detected in about half
of the sample. We discuss the implications of these results on future
projects including the International Pulsar Timing Array (IPTA) and a
PTA based on the Square Kilometre Array. We also present an ``extended
PPTA'' data set that combines PPTA data with earlier Parkes timing
data for these pulsars.



\medskip
\medskip
\end{minipage}
\end{changemargin}
]
\small
\section{Introduction}\label{sec:intro}
Pulsars have many intriguing properties, but their most important
attribute by far is the remarkable stability of the basic pulse
periodicity. Having these ``celestial clocks'' distributed throughout
the Galaxy (with a few in our nearest neighbour galaxies, the
Magellanic Clouds), many of them members of binary systems, makes
possible a range of interesting and important applications. The best
known of these is the detection of orbital decay in the original
binary pulsar, PSR B1913+16 \citep{ht75a}, which provided the first
observational evidence for the existence of gravitational waves (GWs)
and showed that the rate of energy loss was in accordance with the
predictions of Einstein's general theory of relativity (GR)
\citep{wt05}. But there are many others. For example, the pulse
dispersion due to free electrons along the path to the pulsar can
easily be measured and used to study the distribution of ionised gas
in our Galaxy and, potentially, the intergalactic medium. Precise
positions, proper motions and even the annual parallax of pulsars can
be measured using pulsar timing. Careful study of the timing of binary pulsars has
revealed a range of orbital perturbations which not only give
important information about the formation and evolution of the binary
systems, but also allow sensitive tests of gravitational theories. In
particular, the discovery and subsequent timing observations of the
first-known double-pulsar system, PSR J0737$-$3039A/B
\citep{bdp+03,lbk+04}, has allowed four independent tests of GR and
given the most precise verification so far of GR in the strong-field
regime \citep{ksm+06}.

One of the major goals of current astrophysics is the direct detection
of GWs. Detection and study of these waves would open up a new window
on the early Universe and the physics of extreme gravitational
interactions. Enormous effort is going into the construction of
systems such as the Laser Interferometer Gravitational Wave
Observatory (LIGO) \citep{aad+92} and Virgo \citep{aaa+04g} which are
sensitive to GWs with frequencies in the range 10 -- 500 Hz. Initial
versions of these systems are now operating and have placed limits on
the amplitude of GWs from several types of astrophysical source
\citep[e.g.,][]{aaa+06}. Systems with improved sensitivity are being
developed. Advanced LIGO is due for completion in 2014. KAGRA, a 3-km
underground detector with cryogenically-cooled optical systems is
being constructed in Japan \citep{som11} and the Einstein Telescope, a
third-generation detection system, is under development
\citep{paa+10}. The Laser Interferometer Space Observatory (LISA) has
now evolved into eLISA\footnote{See http://www.elisa-ngo.org} which
will be sensitive to GWs with frequencies in the range 0.1 - 100
mHz. Finally, space observatories exploring the cosmic microwave
background such as the Wilkinson Microwave Anisotropy Probe (WMAP)
\citep{bbh+03} and the Planck Surveyor \citep{tmp+10} have as one of
their goals the detection of the B-mode polarisation signature of
primordial GWs; these have a local frequency of $\sim 10^{-9}$~Hz
\citep{cct95}. So far only upper limits have been obtained
\citep[e.g.,][]{ldh+11}.

Because of the great intrinsic stability of their pulsational periods
($P$) or pulse frequencies ($\nu = 1/P$), precision timing
observations of pulsars and, in particular, millisecond pulsars
(MSPs), can in principle be used to detect GWs propagating in our
Galaxy \citep{saz78,det79}. Observed pulse frequencies are modulated
by GWs passing over the pulsar and the Earth; the net effect is the
difference between these two modulations. Pulsar timing analyses give
the differences, commonly known as timing residuals, between observed
pulse times of arrival (ToAs), normally referred to the barycentre of
the solar system, and the predictions of a model for the pulsar
properties \citep[see, e.g.,][]{ehm06}. Pulsar timing is therefore
essentially a phase measurement and is sensitive to the integrated
effect of pulse-frequency modulations. Consequently, it is most
sensitive to long-period modulations out to roughly the data
span. This is typically many years, corresponding to frequencies in
the range 1 -- 30 nHz. Pulsar timing experiments are therefore
complementary to the ground-based and space-based laser interferometer
systems. Since the intrinsic value of $\nu$ and its rate of change,
$\dot\nu$, are a priori unknown, they must be solved for in the timing
solution. Therefore, any external modulation which affects only these
parameters is undetectable.\footnote{An exception to this is where the
  effect is sufficiently large to reverse the sign of $\dot\nu$; this
  is observed for some MSPs accelerated in the gravitational potential
  of globular clusters.}

Different astrophysical sources are likely to dominate the GW spectrum
in the different bands. For example, the GW sources most likely to be
detected by ground-based interferometer systems are the final stages
of coalescence of double-neutron-star binary systems. For eLISA the
most promising source is a similar coalescence of super-massive binary
black holes in the cores of distant galaxies. At an earlier stage of
their evolution, these same super-massive black-hole binary systems
generate a stochastic background of low-frequency GWs. The expected
spectrum of this GW background can be described by a power-law
relation \citep{jhv+06}:
\begin{equation}\label{eq:gwb}
h_{\rm c}(f) = A_{\rm g} \left(\frac{f}{f_{\rm 1 yr}}\right)^\alpha
\end{equation}
where $h_{\rm c}$ is the characteristic GW strain over a logarithmic
frequency interval at frequency $f$ and $f_{\rm 1 yr}$ is
$1/{\rm 1 yr}$.  Predictions for the rate of super-massive black-hole
binary coalescence events following galaxy mergers result in values of
the GW background amplitude $A_{\rm g}$ that are potentially
detectable by pulsar timing experiments
\citep[e.g.,][]{jb03,wl03a,en07,svc08}. For such a GW background, the
exponent $\alpha$ is expected to be $-2/3$ \citep{phi01}. Other
potentially detectable sources include cosmic strings in the early
universe \citep{dv05,smc07} and fluctuations in the primordial
Universe \citep{gri05,bb07}. For these sources, the strain spectrum is
expected to be somewhat steeper with $\alpha \sim -1$.

Observed pulsar periods are affected by many other factors. Pulsars in
binary systems of course have their apparent period modulated by their
binary motion and by relativistic effects. Intrinsic pulsar periods
are not perfectly stable, with two main types of irregularity being
observed: apparently random period ``noise'' and sudden jumps in the
period known as glitches. Fortunately, the strength of these
irregularities is correlated with $|\dot\nu|$ \citep{antt94,wmp+00,sc10}
and so they are weak or undetectable in MSPs as these have very low
values of $|\dot\nu|$ \citep{hlk10}. Pulsar signals suffer
frequency-dependent propagation delays in the interstellar medium,
mainly due to dispersion and usually parameterised by the
dispersion measure (DM), computed using
\begin{equation}\label{eq:dm}
{\rm DM} = K \frac{t_2 - t_1}{f_2^{-2} - f_1^{-2}}
\end{equation}
where $K \equiv 2.410\times 10^{-16}$~Hz$^{-2}$~cm$^{-3}$~pc~s$^{-1}$
and $t_1$, $t_2$ are pulse ToAs measured at radio frequencies $f_1$,
$f_2$ respectively. Because of the changing path to the pulsar
resulting from the motion of the pulsar and the Earth in the Galaxy,
these delays are time-variable and must be continually monitored
\citep{yhc+07}. Errors in the reference atomic timescale and in the
ephemeris used for correction to the solar-system barycentre will
directly affect the observed timing residuals. Finally, the data are
affected by a variety of instrumental effects such as receiver
non-linearities, calibration errors and radio-frequency interference.

These often unknown contributions to the timing residuals make it
effectively impossible to detect GWs with observations of just a few
pulsars. However, data from just one or two pulsars can be used to
place a limit on the GW strength. As an example, \citet{ktr94} used
Arecibo observations of PSR B1855+09 and PSR B1937+21 over an
eight-year data span to place an upper limit of about $6\times
10^{-8}$ on the energy density of a stochastic GW background relative to
the closure density of the Universe $\Omega_{\rm gw}$ at frequencies
around 5 nHz.\footnote{The relative GW energy density is given by
  $\Omega_{\rm gw}(f) = \frac{2\pi^2}{3H_0^2} \, f^2 \, h_{\rm
    c}(f)^2$ \citep{phi01,jhv+06}}

For a positive detection of a stochastic background of GWs, its
effects on the observed timing residuals must be separated from other
contributions to those residuals. Fortunately, a method exists to
achieve this separation. This method depends on making
quasi-simultaneous observations of a number of pulsars that are
distributed across the celestial sphere, thereby forming a ``Pulsar
Timing Array'' (PTA).

For an isotropic stochastic GW background, the expected correlation
between residuals for pairs of pulsars in different directions is a
function only of the angle between the two pulsars \citep{hd83}. The
precise dependence of the correlation on the separation angle depends
on the properties of the GWs themselves \citep{ljp08}.  According to
standard GR, GWs are spin-2 transverse-traceless waves. The
correlation curve for such waves has a maximum of 0.5 for pulsars
which are close together on the sky (reduced from 1.0 because the same
GW background passing over the pulsars produces a statistically equal
but uncorrelated modulation in their residuals), goes negative for
pulsars separated by about $90\degr$ and positive again for $180\degr$
separation (see, e.g., Hobbs et al. 2009a). This well-defined
``quadrupolar'' signature may be compared with the dipolar signature
resulting from an error in the solar-system ephemeris, which
effectively is an error in the assumed position of the Earth relative
to the solar-system barycentre, or the monopolar signature resulting
from time-standard errors which affect all pulsars equally. Other
errors, for example, in the assumed parameters of a binary system, can
be separated by their dependence on binary orbital phase. Errors in
interstellar corrections can be separated by their dependence on
observing frequency. Careful attention to other corrections, for
example, timescale transformations, can ensure that their uncertainty
is negligible \citep{ehm06}.

The idea of observing a set of widely distributed pulsars, both to
reduce statistical uncertainties and to enable the separation of
timing perturbations as described above, was introduced by
\citet{hd83} and further discussed by \citet{rom89} (who coined the
term ``Pulsar Timing Array'') and \citet{fb90}. 

The Parkes Pulsar Timing Array (PPTA) is an implementation of the PTA
concept. It builds on earlier MSP timing efforts at the Parkes 64-m
radio telescope \citep[e.g.,][]{jlh+93,tsb+99,vbb+01}. The PPTA
project is a collaborative effort, primarily between groups at the
Commonwealth Scientific and Industrial Research Organisation (CSIRO)
Astronomy and Space Science Division (CASS) and the Swinburne
University of Technology with important contributions from astronomers
at other institutions, both within Australia and internationally. It
uses the Parkes 64-m radio telescope to observe a sample of 20 MSPs at
intervals of two to three weeks at three radio bands: 50~cm (centre
radio frequency $f_{\rm c} \sim 700$~MHz), 20~cm ($\sim 1400$~MHz) and
10~cm ($\sim 3100$~MHz). As well as the observational program, the
PPTA project has supported work on the analysis of pulsar timing data
and its interpretation. A major component of this was the development
of the {\sc Tempo2} pulsar timing analysis package, described by
\citet{hem06}, \citet{ehm06} and \citet{hjl+08}. The PPTA has also
supported development of the {\sc Psrchive} \citep{hvm04} and {\sc
  Dspsr} \citep{vb11} pulsar data analysis packages.

Detection of the GW background by a PTA was discussed by
\citet{jhlm05}. They showed that precision timing of at least 20
pulsars with ToA errors $\sim 100$~ns and a five-year data span was
necessary for a positive detection of the expected stochastic GW
background from binary super-massive black holes in distant
galaxies. This work formed the basis for the design of the PPTA
observational program. \citet{jhv+06} developed a method of analysing
PTA data sets to place a limit on the energy density $\Omega_{\rm gw}$
of the stochastic background based on the assumption of ``white'' or
uncorrelated timing residuals. They applied this to the \citet{ktr94}
Arecibo data for PSR B1855+09 and Parkes observations of seven PPTA
pulsars from early 2003 to mid-2006 to derive a limit on $\Omega_{\rm
  gw}$ at a frequency of 1/(8~yr) of $1.9\times 10^{-8}$, considerably
better than the limit derived by \citet{ktr94}. \citet{yhj+10}
discussed the sensitivity of a PTA to an isolated GW source, deriving
the first realistic sensitivity curve for a PTA system. Using the
method developed by \citet{wjy+11}, \citet{yhj+10} also placed a
sky-averaged constraint on the merger rate of black-hole binary
systems with a chirp mass\footnote{The chirp mass $M_c =
  (M_1M_2)^{3/5}(M_1 + M_2)^{-1/5}$, where $M_1$ and $M_2$ are the
  masses of the two components of a binary system, is the mass
  parameter relevant to emission of GWs from binary systems
  \citep[see, e.g.,][]{tho87}} of $10^{10}$~M$_\odot$ in galaxies with
redshift $z<2$ of less than one every 5 yr \citep[see also][]{yar11}.
A method of detecting the stochastic GW background based on the
Hellings and Downs correlation and considering the effects of
irregular time sampling and spectral leakage on the detection
sensitivity was developed by \citet{ych+11}. This was applied to the
Parkes observations of \citet{vbv+08} and \citet{vbc+09} showing that
these data were consistent with a null result to 76\% confidence.

PTA observations of sufficient duration can in principle detect
currently unknown solar-system objects, for example trans-Neptunian
objects (TNOs). To date there has been no attempt to directly solve
for an general dipole signature since the unknown dipole axis makes a
general solution difficult. However it is possible to test for
predictable signatures such as errors in the mass of the solar-system
planets. \citet{chm+10} used this method with the additional
assumption that a change in the mass of a planet simply shifted the
barycentre proportionally along the barycentre-planet axis, thereby
introducing a sinusoidal variation at the planetary orbital period
into the observed residuals. They showed that the derived mass of the
Jupiter system was consistent with the value assumed in DE421
\citep{fwb08} to within $2\times 10^{-10}$~M$_\odot$, a higher
precision than that obtained from observations of the {\em Pioneer} and
{\em Voyager} spacecraft and consistent with the value obtained from
{\em Galileo}.

All ToAs are initially measured with reference to a local observatory
clock. They are transferred to an international standard of time, for
example, International Atomic Time TT(TAI), or the annual update of
this, TT(BIPMxx), using clock offsets published by the
BIPM.\footnote{See http://www.bipm.org} Many authors have discussed
the idea of establishing a ``pulsar timescale'' based on the rotation
or orbital motion of pulsars
\citep[e.g.,][]{tay91,pt96,ikr98,rod08}. A pulsar timescale is
fundamentally different to timescales based on atomic frequency
standards in that it is based on different physics -- rotation of
massive bodies -- and is largely independent of Earth-based effects,
for example, seasonal variations. Another important point is that MSPs
will continue to spin for billions of years whereas individual atomic
clocks have lifetimes measured in years or decades at best. Pulsars
may therefore provide a uniquely stable long-term standard of time. A
pulsar timescale is essentially independent of the reference atomic
timescale but it is not absolute and cannot measure linear frequency
drifts of the reference atomic timescale.  However, variations of
higher order can in principle be detected by comparison with a pulsar
timescale. Two recent efforts at establishing a pulsar timescale are
by \citet{rod11} and \citet{hcm+12}. \citet{rod11} used seven years of
archival timing data for six pulsars from the Kalyazin Radio Astronomy
Observatory and an analysis method based on Wiener filtering to
establish a pulsar timescale with stability $\sim 5\times 10^{-14}$
over the 7-year interval. \citet{hcm+12} used Parkes timing data for
19 pulsars spanning up to 17 years, including nearly six years of PPTA
data, to establish the pulsar timescale TT(PPTA11). This had a
stability more than an order of magnitude better than that of
\citet{rod11} and showed significant departures from TT(TAI). These
deviations closely matched the differences between TT(BIPM11) and
TT(TAI) over the same time intervals, thus demonstrating both that
pulsar timescales can be of comparable precision to the best atomic
timescales and that the post-corrections used to form TT(BIPM11) do
improve the stability of the timescale.

PTA data sets have many other applications including detailed studies
of the individual pulsars and studies of the interstellar
medium. \citet{yhc+07} used PPTA data sets to
study variations in interstellar dispersion. \citet{yhc+07a}
and \citet{ychm12} studied the electron density and magnetic field in
the solar wind using observations of several PPTA pulsars that have
low ecliptic latitudes. Detailed studies of the 20cm polarisation
and mean pulse profiles of the PPTA pulsars were presented by
\citet{ymv+11} and \citet{ymh+11} investigated rotation measure
variations for the PPTA pulsars, both short-term variations due to
changes in the Earth's ionosphere and long-term interstellar
variations. \citet{ovh+11} investigated the ultimate limits to
precision pulsar timing in the case of high signal-to-noise ratio
(S/N) observations using J0437$-$4715 as an example.

This paper gives an overall description of the PPTA project which
formally commenced in 2003 June. Preliminary reports on the project
were published by \citet{hob05}, \citet{man06}, \citet{man08},
\citet{hbb+09} and \citet{vbb+10}.  In this paper, the receiver
instrumentation and signal processing systems used and (in some cases)
developed for the PPTA project are described in \S\ref{sec:instr}. The
observational strategy and details of the observed MSP sample are
described in \S\ref{sec:obs} and the results obtained so far are
described in \S\ref{sec:results}.  Prospects for international
collaborations with other PTA projects are discussed in
\S\ref{sec:future} along with future prospects for PTA projects,
especially in the era of the Square Kilometre Array (SKA). The
extended PPTA data set, formed by adding earlier Parkes timing data
for the PPTA pulsars \citep{vbv+08,vbc+09} to the PPTA data set is
described in Appendix A.

\section{Instrumentation and Signal Processing}\label{sec:instr}
\subsection{Telescope and receiver systems}\label{ssec:rcvr}
\begin{figure*}[h]
\centerline{\psfig{file=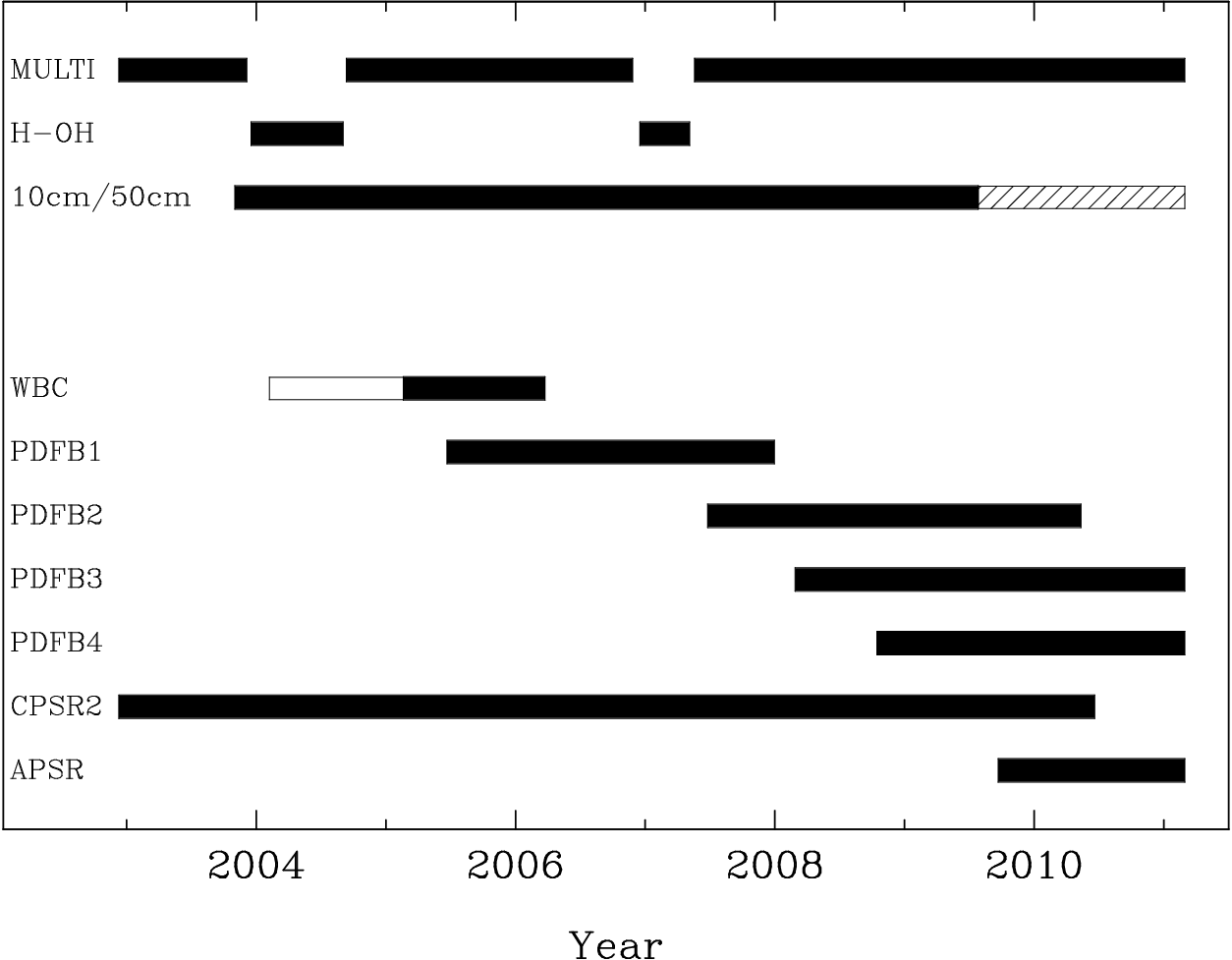,angle=270,width=100mm}}
\caption{Availability of the receivers and backend systems used for
  the PPTA project. The hatched area on the 10cm/50cm bar indicates
  when the 50cm was retuned to operate in the 700 - 770 MHz (40cm)
  band. The WBC bar is open when various instrumental problems
  affected the data quality. Significant intervals of overlap between
  operation of the various backend instruments allowed checks on
  instrument-dependent delays. 
}\label{fg:rcvrs}
\end{figure*}

The PPTA project is based on observations made using the Parkes 64-m
radio telescope in New South Wales, Australia, with a number of
receiver and backend systems.\footnote{We use the term ``receiver'' to
  mean all components of the receiving system from the feed to the
  digitiser inputs and ``backend'' to mean all components from
  the digitiser inputs to the data disks used for near-real-time
  storage of data.}  Parkes is at a latitude of $-33\degr$ and so the
Galactic Centre passes within a few degrees of the zenith. With its
zenith-angle limit of about $59\degr$, it has an effective northern
declination limit of about $+25\degr$, and so can see more than
two-thirds of the celestial sphere, including all of the rich inner
Galaxy. As mentioned in \S\ref{sec:intro}, observations are made in
three different radio-frequency bands : 10cm, 20cm and 50cm. At 20cm,
observations are mostly made using the centre beam of the Parkes
multibeam receiver \citep{swb+96} but occasionally with the ``H-OH''
receiver when the multibeam receiver is not available.  Observations
at 10cm and 50cm are made using the dual-band coaxial ``10cm/50cm''
receiver \citep{gzf+05}. Details of these receivers are given in
Table~\ref{tb:rcvr}, where $f_c$ is the nominal band centre and
$S_{\rm sys}$ is the system equivalent flux density. The original 50cm
band suffered from the presence of digital television (DTV) signals
transmitted from Mt Ulandra, about 200 km south of Parkes. By 2009,
the number of DTV channels in the band had increased to five, covering
more than half the band, and it was decided to move the band up in
frequency to 700 -- 770 MHz (labelled ``40cm''). For convenience, the
band label ``50cm'' implies both 40cm and 50cm data in the remainder
of this paper and we refer to ``three-band'' data sets which in fact
include data with four band designations. The availability of the
various receiver packages is shown graphically in
Figure~\ref{fg:rcvrs}.

All systems receive orthogonal linear polarisations and (except for
the 50cm receiver) have a calibration probe in the feed at $45\degr$
to the two signal probes. A pulsed broad-band noise source is used to
inject a linearly polarized calibration signal which is used to
calibrate the gain and differential phase of the two signal
paths. Because of the coaxial nature of the 10/50cm feed, the 50cm
receiver has four signal probes with opposite signals being added in
$180\degr$ hybrids to form the two orthogonal polarisation
channels. The calibration signal is injected into directional couplers
located between the hybrids and the preamplifiers. Coupling between
the (nominally) orthogonal feeds is low ($<-25$ db) for all receivers
except the Multibeam receiver, where it is about $-12$ db. The
spectrum of the injected calibration signals and the system equivalent
flux density are calibrated in flux density units using
observations of the strong radio source Hydra A (3C 218) as is described
further below (\S\ref{sec:offline}).

\begin{table*}
\begin{center}
\caption{Receivers used for the PPTA project}\label{tb:rcvr}
\begin{tabular}{lcccc}
\hline 
Receiver & Band & Freq. Range & $f_{\rm c}$ & $S_{\rm sys}$ \\
  &  &  (MHz) &  (MHz) & (Jy) \\
\hline 
10/50cm    & 50cm$^*$ & 650 -- 720 & 685  & 56 \\
10/50cm    & 40cm$^\dagger$ & 700 -- 770 & 732  & 62 \\
Multibeam & 20cm & 1230 -- 1530 & 1369 & 36 \\
H-OH      & 20cm & 1200 -- 1800 & 1433 &  30 \\
10/50cm    & 10cm & 2600 -- 3600 & 3100 & 50 \\
\hline
$^*$ Before 2009 July ~~~~~~~$^\dagger$ After 2009 July
\end{tabular}
\end{center}
\end{table*}

For the 20cm and 10cm systems, the signals are down-converted to
baseband in the focus cabin. After transmission to the receiver
control room, the signals from all receivers are down-converted,
band-limited, amplified and adjusted to the optimal signal level in a
remotely controllable down-conversion system \citep{gbs00a} for
presentation to the backend digitiser systems.

Observations are made under control of the Parkes Telescope Control
System (TCS), a graphical interface which allows control of the
telescope pointing and selection of the required receiver, backend
system(s) and observation modes. PPTA observations are normally made
under schedule control ensuring the correct selection of backend
configuration and sequence of calibration and pulsar observations.

\subsection{On-line signal-processing systems}\label{ssec:online}
A number of backend systems have been used in the course of the
project. Their basic parameters are listed in Table~\ref{tb:be}. The
pulsar Wide Band Correlator (WBC) was a correlation spectrometer with
an implementation of the Fast Fourier Transform (FFT) algorithm on
Canaris processor chips and 2-bit (3-level) digitization. The Parkes
Digital FilterBank systems (PDFB1, PDFB2, PDFB3 and PDFB4) are (or
were) implementations of polyphase transforms using Field-Programmable
Gate Array (FPGA) processors with 8-bit digitization of the input
signals \citep[see][for an example implementation and additional
references]{fs04}. Polyphase filterbanks can be designed to have much
superior channel bandpass characteristics compared to FFT
spectrometers. For example, in our systems, the first sidelobes are
$-30$ db down, compared to $-12$ db for an FFT-based
system. Furthermore, the channel bandpass is much more
rectangular. These properties lead to much improved isolation of
narrow-band radio-frequency interference (RFI) and reduced signal
loss. CPSR2 is a baseband recording system allowing coherent
dedispersion of two 64-MHz-wide dual-polarisation data channels and
APSR is its successor allowing coherent dedispersion of
dual-polarisation data at bandwidths of up to 1 GHz.

\begin{figure*}[h]
\centerline{\psfig{file=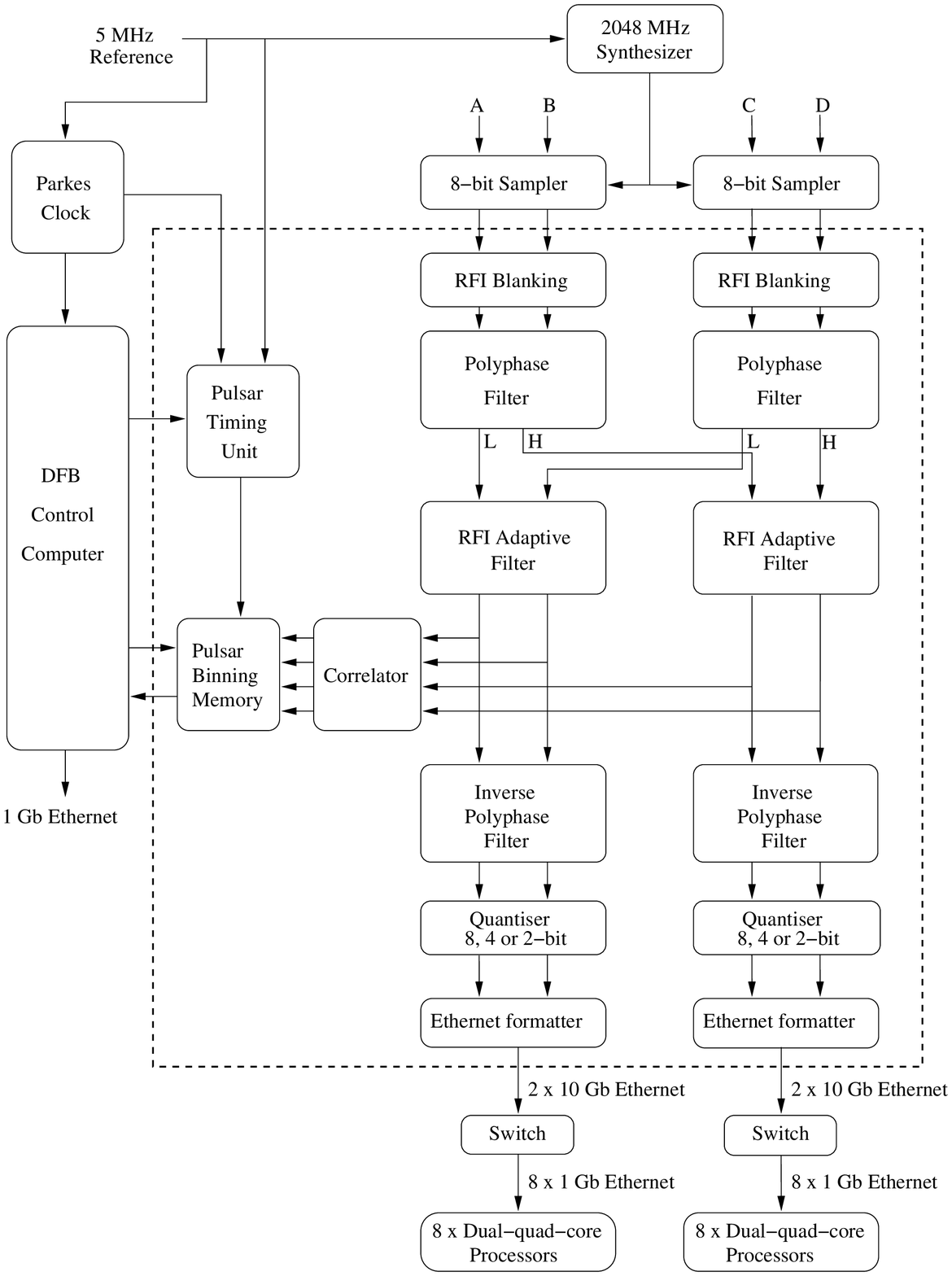,width=125mm}}
\caption{Block diagram of the PDFB3/APSR system. The dashed line
  encloses components contained on the two main DFB boards. The 32-MHz
  synthesizer and samplers are on a separate board. In normal
  operation, the A and B inputs are used for the two
  polarisation channels from the receiver/IF system. The C and D
  channels may be used for the RFI adaptive filter reference input or
  independently for other signals. The L and H
  channels from the polyphase filterbank refer to the lower and upper
  halves of the total bandwidth. Profiles from the pulsar binning
  memory are transferred to the control computer each DFB cycle. The
  APSR baseband outputs are output on two pairs of 10 Gb ethernet
  lines to switches which then distribute the signals amongst the 16
  dual quad-core processors for quasi-real-time dedispersion and
  folding. The Control Computer has control links to most functional
  elements in the system, but most of these are omitted for clarity.
}\label{fg:pdfb}
\end{figure*}

Regular observations with the WBC began in early 2004, but because of
various instrumental problems, high-quality data were not obtained
until early 2005. The WBC was decommissioned in 2006 March. PDFB1 was
based on a commercial (Nallatech) signal-processing board and was in
use from 2005 June to 2007 December. PDFB2 was the first Parkes
digital filterbank system to use processor boards developed at CASS,
namely prototype boards for the Compact Array Broadband (CABB) system
\citep{wfa+11}.  PDFB2 was commissioned with 1024-MHz bandwidth in
June 2007 and decommissioned in 2010 May. PDFB3 and PDFB4 use the
final version of the CABB boards which have more powerful FPGA
processors and have more on-board memory compared to those used for
PDFB2. PDFB3 and PDFB4 were commissioned in 2008 July and remain in
regular use. PDFB3 has two identical signal processing boards allowing
simultaneous processing of up to four input signals and giving better
performance in some configurations. Figure~\ref{fg:pdfb} shows a block
diagram of the PDFB3 system. PDFB4 is similar except that it has just
one processor board and so cannot provide the inverse filter and
subsequent processing for the APSR system (described
below). Figure~\ref{fg:rcvrs} shows the intervals when the various
backend systems were in use for the PPTA project.

\begin{table*}[h]
\begin{center}
\caption{Backend systems used for the PPTA project}\label{tb:be}
\begin{tabular}{lccccc}
\hline 
System & Bandwidth & Dig. Bits & Nr Channels & Nr Bins & $P_{\rm min}$ \\
  & (MHz) & & & & (ms) \\
\hline 
WBC       & 1024 &  2 & 1024  & 1024 & 57 \\
PDFB1     & 256  &  8 & 2048  & 2048  & 83 \\
PDFB2     & 1024 &  8 & 1024  & 2048  & 4.1 \\
PDFB3     & $2\times 1024$ & 8 & 2048 & 2048 & 4.1 \\
PDFB4     & 1024 & 8 & 2048 & 2048 & 4.1 \\
CPSR2     & $2\times 64$  & 2 & -- & -- & -- \\
APSR      & $16\times 64$ & 2 -- 8 & -- & -- & -- \\
\hline
\end{tabular}
\end{center}
\end{table*}

The first five systems in Table~\ref{tb:be} provide real-time
computation of both the direct and cross products of the two complex
baseband signals (A,B) giving full polarisation information, i.e.,
A*A, B*B, Re(A*B) and Im(A*B). They also provide real-time folding of
the channel and polarisation product data at the apparent pulsar
period with the specified maximum number of bins across the pulse
period. All systems are (or were in the case of the first three)
controlled by a computer which obtains control data from TCS,
including telescope, source and configuration data needed to
sychronise the observations. The control computer operates on a basic
cycle which is normally of 10 s duration. It provides displays of
receiver bandpasses, digitiser levels and folded pulse profiles each
cycle.

The bandwidths given in Table~\ref{tb:be} are maximum values; all
bandwidths less than the maximum by factors of two down to 8 MHz can
be processed provided an appropriately band-limited signal is
presented at the digitiser inputs. Observed bandwidths are normally
64~MHz for the 50cm receiver, 256~MHz for the 20cm receivers and
1024~MHz for the 10cm receiver. The number of frequency channels and
phase bins quoted in Table~\ref{tb:be} are also maximum values in the
sense that their product cannot exceed the product of the quoted
values; the number of channels and bins can be varied by factors of
two within this limitation. Qualifications on this are that for the
WBC, the quoted number of channels is an absolute maximum and for
PDFB1 the number of channels could be changed downward only by factors
of four. For PDFB2, PDFB3 and PDFB4, the maximum number of channels
per polarisation product is 8192. In Table~\ref{tb:be} the minimum
folding period $P_{\rm min}$ is given for the specified bandwidth,
number of channels and number of bins. $P_{\rm min}$ is (down to some
limiting value) inversely proportional to the bandwidth and directly
proportional to the channel--bin product. For example, for a bandwidth
of 256 MHz, 1024 bins and 1024 channels, $P_{\rm min}$ for PDFB3 and
PDFB4 remains at 4.1 ms, but for 1024 MHz bandwidth, 512 bins and 512
channels, it is just 0.26 ms.

The WBC and PDFB1 systems used {\sc Tempo} polyco files as a basis for
predicting the topocentric folding period; PDFB2, PDFB3 and PDFB4
use(d) {\sc Tempo2} prediction files which have greater precision
\citep{hem06}. Pulsar astrometric parameters and barycentric period
and binary data are obtained using {\sc psrcat} \citep{mhth05} with a
regularly updated database file.\footnote{See
  http://www.atnf.csiro.au/research/pulsar/psrcat} An on-board
``Pulsar Timing Unit'' (PTU) increments the bin number at the rate
required to maintain phase with the apparent pulse period, the latter
being updated each cycle of the control computer, normally at
10-second intervals. The PTU and the input 8-bit digitisers use a 5
MHz signal locked to the Observatory H-maser as a primary reference;
this is multiplied to 256 MHz and phase-locked to the Observatory
1-second (1 PPS) signal for the PTU and then further multiplied to
2048 MHz for the digitisers that Nyquist-sample the 1024 MHz input
bands. The PTU determines the lag between the zero of pulse phase at
the start of an observation and the leading edge of the 1 PPS signal
with an uncertainty of 4~ns and sets the topocentric folding period or
bin time each cycle with a precision typically better than $1:10^9$.

Folded profiles for the four polarisation products for every frequency
channel are transferred to the control computer every cycle. They are
integrated there for a ``sub-integration'' time, which is a multiple
of the cycle time and normally 60~s, before being written to disk
along with header information in {\sc psrfits} format
\citep{hvm04}.\footnote{See also
  http://www.atnf.csiro.au/research/pulsar/psrfits.} The {\sc psrfits}
output files also contain tables giving other information such as mean
digitiser levels as a function of time, the distribution of digitiser
counts, the receiver bandpass for each polarisation channel, the
pulsar parameters used for the prediction of the folding period and
the predictor table used for the observation.

CPSR2 (Caltech-Parkes-Swinburne Recorder, Version 2) was a baseband
recording system which recorded two pairs of 64-MHz-wide baseband
signals with 2-bit (4-level) digitization \citep{bai03,hot07}. Data
were distributed to two primary processors and then to 28 secondary
processors that performed coherent dedispersion of the baseband data
\citep{hr75}, followed by folding at the apparent pulsar period to
form 128-channel pulse profiles, typically with 1024 pulse-phase bins
per polarisation product. CPSR2 was commissioned in August 2002 and
decommissioned in July 2011. CPSR2 recorded data files every 16 s in
``timer'' format. These files were visually checked for RFI and then,
if clean, summed to form 64-s sub-integrations. The sub-integration
files for each observation were then combined with additional header
information to form a {\sc psrfits} file.

APSR (ATNF-Parkes-Swinburne Recorder) is a baseband recording system
which uses PDFB3 for digitisation and signal conditioning
(Figure~\ref{fg:pdfb}). It provides 16 pairs of baseband signals
covering a maximum total bandwidth of 1024 MHz. Because of data
transfer limitations, for 1024 MHz and 512 MHz total bandwidth,
samples are truncated to two and four bits respectively; for smaller
bandwidths, 8-bit data are recorded. Data are transferred via four 10
Gbit ethernet lines and a fast switch to a cluster of 16
dual-quad-core processors where coherent dedispersion and folding are
carried out using {\sc Dspsr} \citep{vb11}. The data are stored on
disk at Parkes and then converted to {\sc psrfits} format for
subsequent processing. APSR has a web-based real-time monitoring and
control system which communicates with TCS to synchronise the data
recording. APSR commenced regular operation in 2009.

PDFB3 has an option for real-time rejection of radio-frequency
interference (RFI) which can be used for both fold-mode and APSR
observations. Two types of rejection are provided: a) time-domain
clipping of impulsive broad-band interference and b) frequency-domain
adaptive filtering of quasi-stationary RFI \citep{khc+05,kmbh10}. The
latter requires a reference signal containing the RFI which is fed to
one of the second pair of digitiser inputs. Provided the RFI in the
reference signal has sufficient S/N, the adaptive
filter removes the RFI from the two polarisation channels without disturbing
the underlying astronomy signal. Its main application was to the
original 50cm band (650 - 720 MHz) where digital television signals
from transmitters on Mount Ulandra, located approximately 200 km south
of Parkes, were at significant levels. The reference signal was
obtained from a 4-m reflector with a vertically polarised dipole feed
(to match the transmitted polarisation) directed at the horizon in the
direction of Mount Ulandra. The same reference signal cleans
both polarisations of the astronomy signal, preserving the pulse
polarisation.

Quasi-real-time monitoring of pulsar profiles as a function of time,
frequency and pulse phase together with input bandpass profiles is
provided for all operating backend systems by a web-based monitoring
system.\footnote{http://pulseatparkes.atnf.csiro.au/dev/}

\subsection{Calibration}
Calibration of the data is important to reduce systematic errors
associated with different bandpass gains and phases for the two
polarisation channels, to place the Stokes parameters in a celestial
reference frame, and to correct for the effects of cross-coupling in
the feed \citep{van04c,van06}. Parameters describing the orientation
of the signal and calibration probes relative to the telescope axis
are stored in the main header of each {\sc psrfits} file
\citep{vmjr10}.  Short (typically 2~min) observations of the pulsed
calibration signal preceding (and sometimes following) each pulsar
observation are used to determine the instrumental gain and phase. The
calibration data are then applied to the pulsar observations using the
{\sc Psrchive}\footnote{See http://psrchive.sourceforge.net} program
{\sc pac} to flatten the bandpass and transform the polarisation
products to Stokes parameters. Parallactic rotation is also corrected
to place the Stokes parameters in the celestial reference frame.

For all systems, input signal levels are automatically adjusted to the
optimal operating point (within $\pm 0.5$~db) as part of the
calibration procedure. For the PDFB systems, the operating point was
chosen to give an rms variation of 10 digitiser counts which ensured
linearity while preserving adequate headroom for strong signals.

The pulsar data are placed on a flux density scale utilising
observations of Hydra A, assumed to have a flux density of
43.1 Jy at 1400 MHz and a spectral index of $-0.91$ over the PPTA
frequency range. These observations are normally made once per session
for all three bands and consist of a sequence of five 2-min
calibration observations at positions off north -- on source -- off
south -- on source -- off north, where the off-source positions are
$1\degr$ from the source position. The data are processed using the
{\sc Psrchive} program {\sc fluxcal} to produce flux calibration files
for each band. These are subsequently used by {\sc pac} to calibrate
the pulse profiles in flux density units. 

A sequence of observations of PSR J0437$-$4715 are made for each
receiver system several times a year to calibrate the feed
cross-coupling. Usually 8 -- 10 observations, each of 16-min duration,
are made across the 10.5 hours that the pulsar is above the telescope
horizon. These data are processed using the {\sc Psrchive} program
{\sc pcm} with the ``Measurement Equation Modelling'' (MEM) method
\citep{van04c} to form ``{\sc pcm}'' files. These can be used by {\sc
  pac} to correct the pulsar data files for the effects of feed
cross-coupling.  For this paper, cross-coupling corrections are only
applied to data obtained using the 20cm multibeam receiver.

\subsection{Off-line Signal Processing}\label{sec:offline}
\begin{figure*}
\begin{minipage}{150mm}
\begin{tabular}{cc}
\mbox{\psfig{file=J1713_YFp.ps,width=70mm,angle=270}} 
      & \mbox{\psfig{file=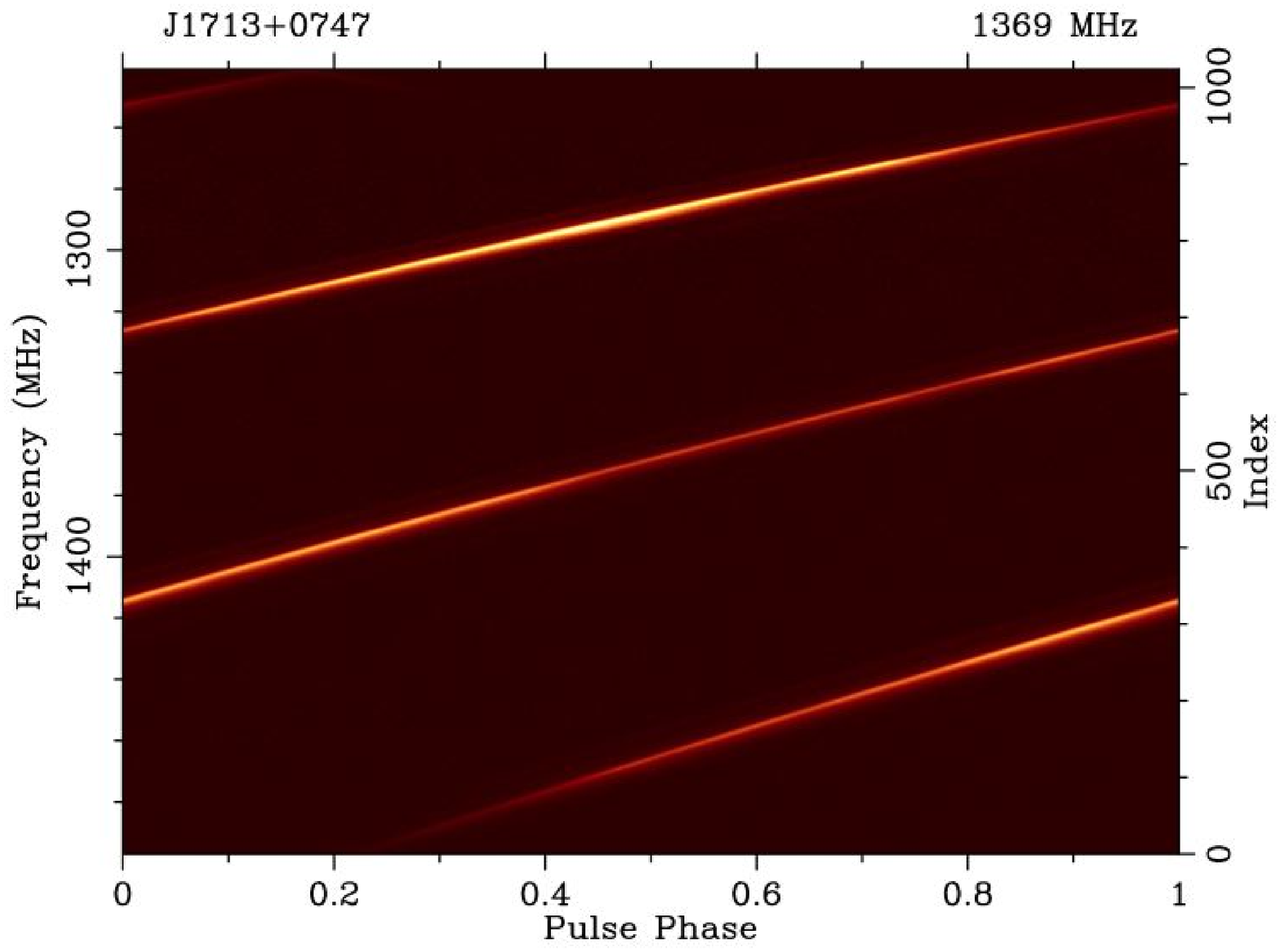,width=70mm}} \\
\mbox{\psfig{file=J1713_prf.ps,width=65mm,angle=270}}
      & \mbox{\psfig{file=J1713_bpass.ps,width=70mm,angle=270}} 
\end{tabular}
\caption{Total intensity (Stokes $I$) pulse profile displays formed
  using {\sc Psrchive} routines for a 20cm PDFB4 observation of PSR
  J1713+0747. The upper-left plot is a false-colour image of the
  dedispersed pulse profile for each 1-min sub-integration, the
  upper-right plot shows the profile summed in time as a function of
  frequency across the band. The lower left plot is the mean pulse
  profile summed in time and frequency and the lower right plot shows
  the receiver bandpass for the two polarisations which are summed to
  form the total intensity. The upper plots show raw uncalibrated data
  whereas data for the mean profile plot has been band-pass and
  flux-density calibrated after excision of the few narrow RFI signals
  visible on the band-pass plot. A decrease in pulse intensity
  resulting from diffractive interstellar scintillation over the
  one-hour observation is visible in the upper-left plot whereas most
  of the frequency-dependent variations seen in the upper-right plot
  result from the instrumental bandpass.}
\label{fg:pulse_plots}
\end{minipage}
\end{figure*}

All manipulation, visualisation and analysis of pulse-profile data is
done using the {\sc Psrchive} pulsar signal processing system. Data
may be calibrated, shifted in pulse phase, summed in time, corrected
for dispersion delays and summed over frequency channels. {\sc
  Psrchive} routines are also used for RFI excision and ToA
estimation. A wide variety of different display formats is available,
displaying the data as functions of pulse phase, time, frequency,
polarisation parameters, etc. Programs to list file header and profile
data in simple ascii formats are also
provided. Figure~\ref{fg:pulse_plots} shows typical displays for PSR
J1713+0747 from a standard 64-min observation at 20cm with PDFB3.

\begin{figure*}[ht]
\centerline{\psfig{file=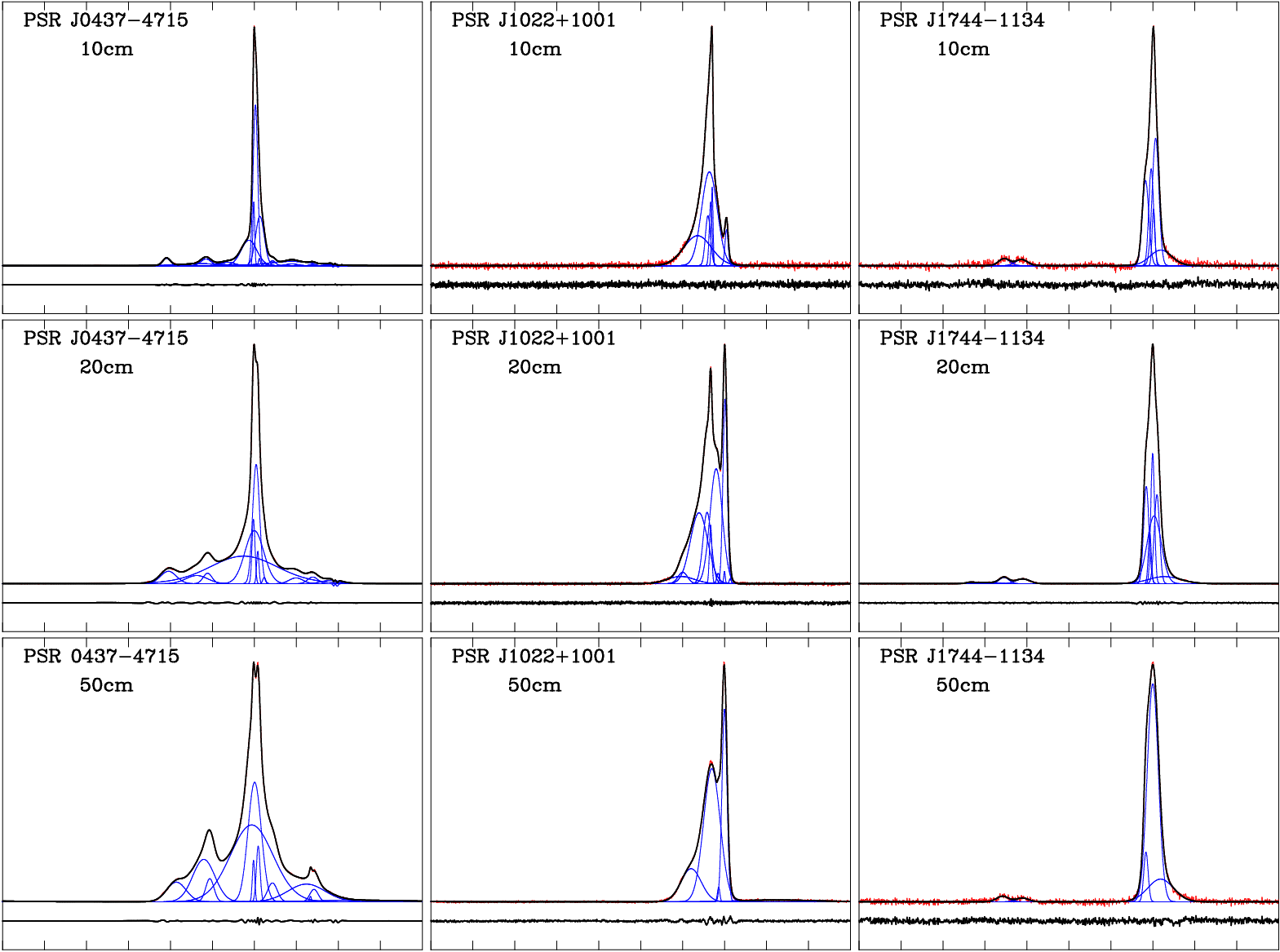,angle=-90,width=\textwidth}}
\caption{Observed mean pulse profiles (red), fitted von Mises
  components (blue), the noise-free template obtained by summing the
  components (black) and, offset below the other profiles, the
  difference between the mean pulse profile and the template profile,
  for three of the PPTA pulsars at each of the three observing
  bands. The full pulse period is shown in each case. For PSR
  J0437$-$4715, the mean pulse profiles are invariant interval; for
  the other two pulsars they are total intensity.}\label{fg:tmplt}
\end{figure*}

Pulse ToAs are important for many research areas including, of course,
the PPTA project. While there are several options within {\sc
  Psrchive}, ToAs and their uncertainties are obtained by performing a
Fourier-domain cross-correlation \citep{tay90a} of the observed pulse
profile with a standard template for each pulsar. Noise-free standard
templates were formed by interactively fitting scaled von Mises functions (using the
{\sc Psrchive} program {\sc paas}) to a high S/N observed profile
formed by adding many individual observations
\citep[cf.,][]{ymv+11}. The number of fitted functions varied
depending on the complexity of the profile and was increased until the
peak residual was less than or about three times the baseline rms
deviation. For all pulsars except PSR J0437$-$4715, the
profiles and template are total intensity or Stokes $I$; for PSR
J0437$-$4715 it is advantageous to fit to the invariant-interval
profile, i.e., the profile of $(I^2 - P^2)^{1/2}$, where $I$ is the
total-intensity Stokes parameter and $P$ is the polarised part of the
signal, $P=(Q^2+U^2+V^2)^{1/2}$, where $Q$, $U$ and $V$ are the Stokes
parameters describing the signal polarisation 
\citep{bri00}. This largely avoids possible systematic errors
associated with polarisation calibration of the
data. Figure~\ref{fg:tmplt} shows mean observed pulse profiles,
component von Mises functions and the template formed by summing these
for three representative PPTA pulsars at each of the three observing
bands. For PSR J0437$-$4715 there are 13, 14 and 17 von Mises
components at 10cm, 20cm and 50cm respectively. Other pulsars have
less than this, but never less than three (for J1909$-$3744 at 10cm
where the profile is narrow and has little structure). The ToA
reference phase is at the highest peak of the 20cm profile. Template
profiles at 10cm and 50cm for a given pulsar were cross-correlated
with the 20cm profile and aligned for maximum correlation.

Each instrument has an effective signal delay (or advance) resulting
from details of the time-tagging mechanism and signal processing
delays. These can be many tens of microseconds and must be calibrated
before data from different instruments are combined or compared. For the PDFB
systems, delays were measured by modulating the system noise with a
PIN-diode in the signal line just before it enters the down-conversion
system. The modulation signal was generated by a programmable pulse
generator and consisted of a pulse train (usually 6 or 12 pulses) with
40\% duty cycle and total duration of about 0.95 ms, the first of
which was triggered by the leading edge of the 1-ms pulse from the
Observatory clock. This trigger is also synchronous with the leading
edge of the 1-sec clock pulse which is used for the PDFB time
tagging. This pulse train was observed with each instrument with a
folding period of exactly $n$~ms where $n$ was typically 2 --
5. Observation times were typically 2~min; several observations were
made for each configuration, often in different sessions.

\begin{figure}[ht]
\centerline{\psfig{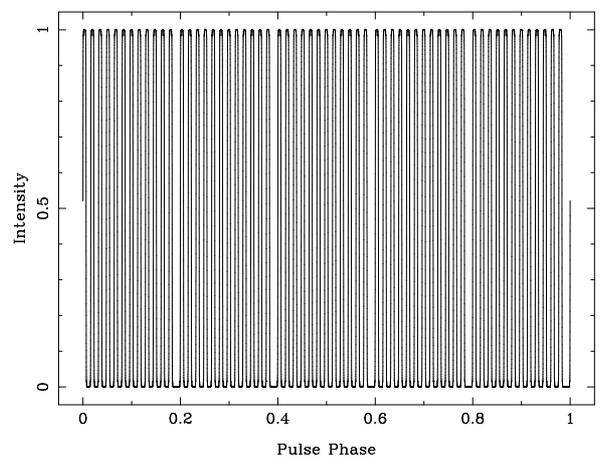}}
\caption{Template profile used for measurement of instrumental delays
  for the PDFB3/4 configuration with 256 MHz bandwidth, 1024 channels
  and 1024 profile bins. The rectangular input waveform has a leading
  edge at phase 0.0; the convolution for the finite channel bandwidth
  smears the edge transitions by a small amount.  }\label{fg:delay}
\end{figure}

A simulated pulse train matching the modulation signal was then
convolved with the impulse response corresponding to the channel width
of the particular PDFB configuration to produce a reference template
for use with the {\sc Psrchive} program {\sc pat} used to produce
ToAs. An example of such a convolved template is given in
Figure~\ref{fg:delay}. The offset of the observed pulse profile
relative to the template profile, modulo 1~ms, given by {\sc pat} is
the instrumental delay for the particular instrument and
configuration. This measurement typically had an uncertainty of a few
ns for a given observation, but additional variable delays of 10 - 200
ns were revealed for some instruments. These variable delays, which
occur apparently randomly and remain to be identified, may contribute
a small amount of additional effectively white noise to observed
ToAs. Additional constant offsets of 609~ns (estimated cable delay
from the focus cabin to the point where the PIN modulator was
inserted), 400~ns (instrumental delay in the pulse generator) and
30~ns (effective propagation delay from the focus cabin to the
intersection of axes of the telescope) were subtracted from the
measured delays to refer derived pulse ToAs to the intersection of the
azimuth and elevation axes (the topocentric reference point) of the
telescope.

Correction for these instrumental delays allowed most of our
observations to be placed on a common timescale. However, for some of
the early instruments and configurations, these measurements were not
made. In these cases, differential offsets from instruments with
measured delays were determined by comparison of simultaneous or
contemporaneous ToAs for several of the more precisely timed
pulsars. For a few of the most precisely timed pulsars, simultaneous
measurments with different instruments revealed systematic ToA
offsets, generally $< 100$~ns, which probably result from a
frequency-dependence of the relevant pulse profile. These were
compensated for by a small phase rotation of the appropriate profile
template.  Measured delays are contained in a file which is
automatically accessed by the PPTA processing pipeline, the operation
of which we discuss in \S\ref{ssec:pipeline}. Occasionally, file
header parameters were incorrect, usually during the commissioning
phase for a new instrument. These header errors are also corrected in
the processing pipeline.

Observed ToAs are intially measured with reference to the Parkes
Observatory hydrogen maser frequency standard, UTC(PKS). These are
referred to one of the international timescales published by the
Bureau International de Poids et Mesures (BIPM), for example, TT(TAI)
or one of its retroactive revisions TT(BIPMxx), where 20xx is the year
up to which the timescale is computed \citep[see, for
example,][]{pet05}. In this paper, we use TT(BIPM11), available at the
BIPM ftp website.\footnote{ftp://tai.bipm.org/TFG/TT(BIPM)} Two
methods of time transfer are currently available. The first is based
on a Tac32Plus Global Positioning System (GPS) clock which directly
gives UTC(GPS) - UTC(PKS) at 5-minute intervals. The BIPM publishes
tables (in Circular T) from which daily values of TT(TAI) - UTC(GPS)
and hence TT(TAI) - UTC(PKS) may be computed. The second system uses a
GPS common-view link to UTC(AUS), operated by the National Measurement
Institute in Sydney, giving UTC(AUS) - UTC(PKS) also at 5-min
intervals. Circular T also gives daily offsets of UTC - UTC(AUS) from
which TT(TAI) - UTC(PKS) may be derived. Daily averages of these clock
corrections are available to {\sc Tempo2}. Typically the two modes of
time transfer differ by a few nanoseconds after removal of a constant
offset.

Observed ToAs must be referred to the solar-system barycentre (SSB),
assumed to be an inertial (unaccelerated) reference frame, before
comparison with predicted arrival times based on a model for the
pulsar. This transformation uses the Jet Propulsion Laboratory DE421
solar-system ephemeris \citep{fwb08} to correct for the motion
of the Earth relative to the SSB, as well as other terms discussed in
detail by \citet{ehm06}. Variations in interstellar and solar System
dispersion also affect observed ToAs and must be corrected
for using measured or estimated dispersion measures to refer the ToAs
to infinite frequency \citep{yhc+07,yhc+07a}. 

The PPTA project involves frequent observations over a long data span,
generating large amounts of data, approximately two terabytes per
year. Immediately after completion of an observation, data files are
automatically transferred to archive disks at CASS Headquarters in
Sydney. As described by \citet{hmm+11}, raw data files for most of
these observations are publically available on the CSIRO Data Access
Portal,\footnote{https://data.csiro.au/dap/} which is part of the
Australian National Data Service,\footnote{http://www.ands.org.au}
after an embargo period of 18 months from the time of observation.

\subsubsection{Processing Pipeline}\label{ssec:pipeline}
As data files are transferred to Sydney, details of the observation
including file name, pulsar name (or calibration identifier),
telescope name, pointing right ascension and declination, receiver,
observing frequency, backend instrument and configuration, bandwidth,
number of channels, number of subintegrations, number of
polarisations, project code and {\sc psrfits} version number are
recorded in a mySQL table. Separate tables are used for
calibration files, pulsar observation files, flux-calibration files,
{\sc pcm}-calibration files and profile template files.  A processing
script is then run to execute the following sequence of commands
for each new calibration or observation file. Relevant {\sc Psrchive}
programs are given in parentheses.
\begin{itemize}
\item For pulsar files, the DM and pulsar parameter table are updated.
  ({\sc pam})
\item The observing band (10cm, 20cm or 50cm) is determined from the
  header data. Frequency ranges known to be contaminated by RFI for a
  given band and observation time are given zero weight. Band edges
  (5\% of the bandwidth) are also given zero weight, mainly to remove
  aliased out-of-band signals. Pulse profile phase bins affected by
  undispersed transient RFI are set to a local mean. ({\sc paz})
\item Calibration files are checked for a bad first sub-integration
  (which occasionally occurs because of faulty synchronisation of the
  observation start); the sub-integration is given zero weight if
  faulty. ({\sc cleanCalFile})
\item For observation files, data are averaged in time to give files with
  eight sub-integrations. Calibration files are fully summed in
  time. ({\sc pam})
\item Header data known to be invalid are corrected. ({\sc fix\_ppta\_data})
\item Observation start times are adjusted to compensate for
  instrumental delays, thereby referring the start time to the
  telescope intersection of axes. ({\sc dlyfix})
\item Profiles are calibrated, correcting for instrumental gain and
  phase and placed on a flux density scale. ({\sc pac})
\item This calibration step is repeated for 20cm data including the
  MEM calibration. ({\sc pac})
\item Profile data are summed to form 32 frequency channels and to form either
  the Stokes parameters or (for PSR J0437$-$4715) the polarisation
  invariant interval. The resulting files are stored in the data archive
  and referenced in the database. ({\sc pam})
\item Profile data are fully summed across frequency and time and
  polarisations combined to form either Stokes $I$ or invariant-interval
  profiles which are stored in the data archive and referenced in the
  database. ({\sc pam})
\item The fully summed profiles are cross-correlated with an
  appropriate template profile to form pulse ToAs which are stored in the
  database ({\sc pat})
\end{itemize}

A set of scripts are available to obtain information from the
database, for example, header data for a given observation or pulse
ToAs for a given pulsar with or without the MEM calibration. The
ToAs are provided in {\sc Tempo2} format and include flags for the
observing band and receiver-backend system. {\sc Tempo2} can
use these flags as well as observation times (in MJD), observation
frequencies or ToA uncertainties to select a particular subset of the
observations using command-line arguments \textsc{-pass} and
\textsc{-filter}. More complex selection can be carried out within a
\textsc{select} file that contains a list of filters. 

Observations that are affected by remaining RFI, calibration or other
instrumental errors are examined and corrected if
possible. If uncorrectable, the affected observation is flagged as bad
in the database and not included in subsequent processing or analysis.

\section{Observational Strategy}\label{sec:obs}
Simulations of the predicted stochastic GW background from binary
supermassive black holes in galaxies \citep{jb03,wl03a,en07,svc08}
show that timing of about 20 pulsars with weekly observations over a
five-year data span with rms residuals of about 100 ns are needed for
a significant detection \citep{jhlm05,hjl+08}. For a background as
described by Equation~\ref{eq:gwb}, the (one-sided) power spectrum of
the timing residuals is a power law given by
\begin{equation}\label{eq:gwb_res}
P_{\rm g}(f) = \frac{A_{\rm g}^2}{12\pi^2}\left(\frac{f}{f_{\rm 1 yr}}\right)^{2\alpha-3}
\end{equation}
\citep[e.g.,][]{jhv+06}. For $\alpha = -2/3$, the spectral exponent of
the residual fluctuations is very steep, $-13/3$. Consequently, the
sensitivity of a PTA is greatly improved with longer data spans as the
GW background signal dominates the overall spectrum at low
frequencies. As will be discussed more fully in \S\ref{sec:future}
below, the minimum detectable GW signal is roughly proportional
to the amplitude of the timing residual fluctuation and therefore
scales approximately as $T^{13/6}$, where $T$ is the data span. It
also scales approximately as $N$, the number of pulsars in the PTA
\citep{vbc+09}. It is important to note that with significantly less
than 20 pulsars, no matter how precise their ToAs, it would be
impossible to make a significant detection of the expected stochastic
GW background. This is primarily a consequence of the GW self-noise, that is,
the random noise introduced into the correlation signature
\citep{hd83,hjl+08} by the uncorrelated GWs passing over each pulsar.

\subsection{Sample Selection}
\begin{figure*}[ht]
\centerline{\psfig{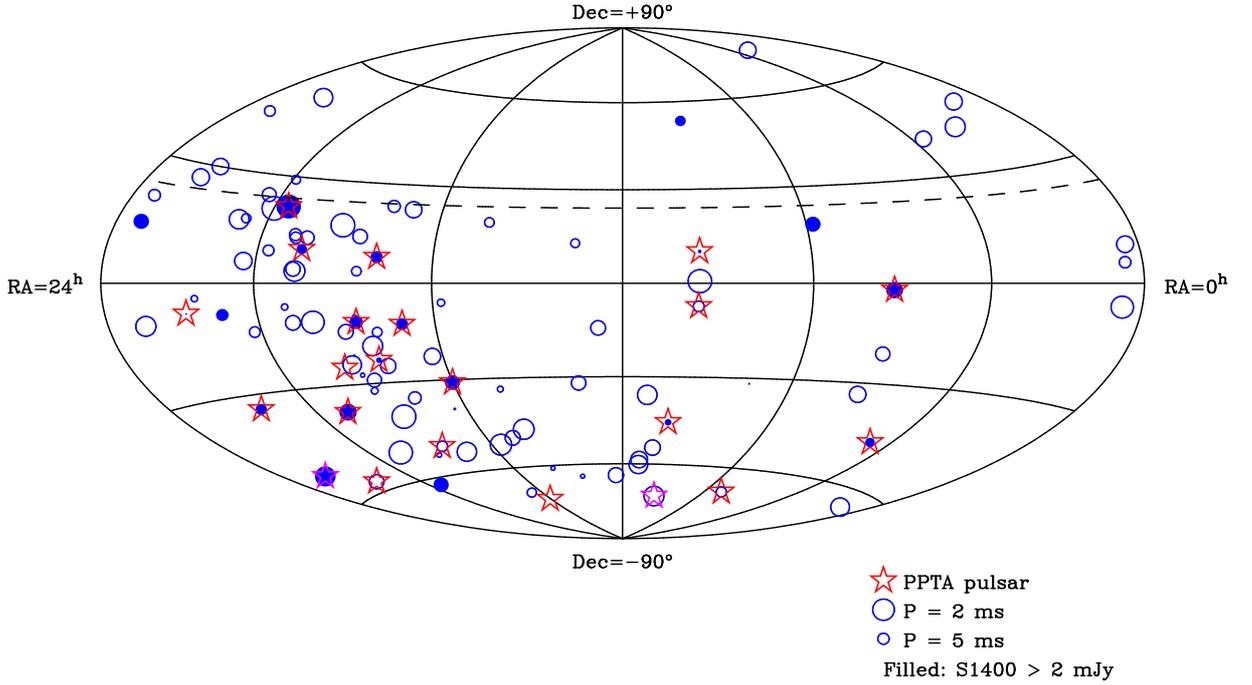}}
\caption{Distribution in celestial coordinates of pulsars suitable for
  pulsar timing array observations. All are radio-emitting MSPs
  (with $P<20$~ms) in the Galactic disk except PSR J1824-2452A
  which is an MSP located in the globular cluster M28 (see text).  The
  area of the plotted circle is inversely proportional to the pulsar
  period and the circles are filled for pulsars with mean flux density
  above 2 mJy. The dashed line is the northern declination limit of
  the Parkes radio telescope. Pulsars selected for the PPTA are marked
  with a star, red for the original 20 pulsars and mauve for the two
  pulsars recently added to the PPTA sample (see text).}\label{fg:ppta_sky}
\end{figure*}

Only MSPs have the period stability and potential ToA precision to be
useful for a PTA. About half of all known MSPs are located in globular
clusters, but these pulsars are less useful partly because they are
often very weak, but more importantly, because gravitational
interactions with other cluster stars introduce additional
perturbations to the observed pulsar period. Similarly, because of the
relativistic perturbations, pulsars in very tight binary orbits are
not ideal, especially if the orbit is eccentric. ToA uncertainties are
approximately proportional to pulse width divided by the S/N.
Consequently for timing array purposes, relatively strong
Galactic-disk MSPs with short periods and/or narrow pulses and either
isolated pulsars or pulsars in wide binary orbits are
preferred. Fig.~\ref{fg:ppta_sky} shows the distribution on the
celestial sphere of MSPs suitable for timing arrays and those selected
for the PPTA.\footnote{Data from the ATNF Pulsar Catalogue V1.44.}
This figure illustrates the fact that at present there are few pulsars
suitable for PTA observations that are inaccessible to the Parkes
telescope.

Table~\ref{tb:ppta_psrs} lists the MSPs selected for the
PPTA. Following the pulsar J2000 name, the pulsar period $P$, DM and
orbital period $P_{\rm b}$ (if applicable) and the standard observation
time are given. The next six columns
give the mean and rms pulse flux densities (averaged over the
observation time) for the 50cm (700 MHz), 20cm (1400 MHz) and
10cm (3100 MHz) bands, respectively. The final two columns give the
mean pulse width at 50\% of the peak level for the 20cm (1400 MHz) total
intensity pulse profiles and its rms uncertainty. These flux densities
and pulse widths are derived from PPTA observations as described in
\S\ref{sec:flux} below. All of these pulsars, with one exception,
are located in the Galactic disk. The exception, PSR J1824-2452A, is
located in the globular cluster M28. It was included in the PPTA
sample partly because it is relatively strong and can be accurately
timed and partly to investigate the effects of period irregularities
on timing-array analyses.  As well as the globular-cluster
perturbations, this pulsar has relatively large DM
variations. Furthermore, a small period glitch was reported for this
pulsar \citep{cb04}. As Fig.~\ref{fg:ppta_sky} shows, the selected
pulsars are widely distributed on the celestial sphere and
consequently provide a good range of angular separations for the
correlation analysis.

\begin{table*}[h]
\begin{center}
\caption{The PPTA pulsars: basic parameters, observation times, flux
  densities and pulse widths}\label{tb:ppta_psrs}
{\footnotesize
\begin{tabular}{ld{4}d{4}d{5}cccccccd{4}d{3}}
\hline 
\multicolumn{1}{c}{PSR} & \multicolumn{1}{c}{$P$} & \multicolumn{1}{c}{DM} &
\multicolumn{1}{c}{$P_{\rm b}$} & \multicolumn{1}{c}{$T_{\rm obs}$} & 
\multicolumn{2}{c}{S$_{700}$} & \multicolumn{2}{c}{S$_{1400}$} & 
\multicolumn{2}{c}{S$_{3100}$} & \multicolumn{2}{c}{W$_{1400}$} \\
 & & & & & Mean & Rms & Mean & Rms &  Mean & Rms &  \multicolumn{1}{c}{Mean} & \multicolumn{1}{c}{Rms} \\
 & \multicolumn{1}{c}{(ms)} & \multicolumn{1}{c}{(cm$^{-3}$ pc)} &
\multicolumn{1}{c}{(d)} & \multicolumn{1}{c}{(min)} & (mJy) & (mJy) &
(mJy) & (mJy) & (mJy) & (mJy) & \multicolumn{1}{c}{(ms)} & \multicolumn{1}{c}{(ms)} 
 \\
\hline 
J0437$-$4715 & 5.757 & 2.65 & 5.74   & 64 & 406 &502 & 149 & 36 & 32.3 &3.2&0.1410&0.0005 \\
J0613$-$0200 & 3.062 & 38.78 & 1.20  & 64 & 7.2 &0.7 & 2.3&0.3 & 0.42&0.64 &0.462&0.001\\
J0711$-$6830 & 5.491 & 18.41 & ...   & 64 & 6.6 &4.5 & 3.2&5.7 & 0.52&0.26 &1.092&0.005\\
J1022+1001   & 16.453 & 10.25 & 7.81 & 64 & 5.7 &2.5 & 6.1&5.4 & 1.30&0.30 &0.972&0.005 \\
J1024$-$0719 & 5.162 & 6.49 & ...    & 64 & 5.4 &4.4 & 1.5&1.1 & 0.37&0.13 &0.521&0.010 \\
J1045$-$4509 & 7.474 & 58.15 & 4.08  & 64 & 9.2 &1.9 &2.7 &0.7 &0.48 &0.10 &0.840&0.012 \\
J1600$-$3053 & 3.598 & 52.19 & 14.34 & 64 & 3.4 &0.4 &2.5 &0.4 &0.77 &0.17 &0.094&0.001\\
J1603$-$7202 & 14.842 & 38.05 & 6.31 & 64 & 12.1&2.9 &3.1 &0.9 &0.32 &0.12 &1.206&0.003\\
J1643$-$1224 & 4.622 & 62.41 &147.02 & 64 & 15.1&0.6 &4.8 &0.4 &1.10 &0.12 &0.314&0.002\\
J1713+0747   & 4.570 & 15.99 & 67.83 & 64 & 8.9 &6.6 &10.2&10.8&2.74 &1.90 &0.110&0.001\\
J1730$-$2304 & 8.123 & 9.61 & ...    & 64 &11.2 &3.6 &3.9 &1.9 &1.97 &2.35 &0.965&0.004\\
J1732$-$5049 & 5.313 & 56.84 & 5.26  & 64 & 6.9 &2.7 &1.7 &0.3 &0.37 &0.06 &0.292&0.003\\
J1744$-$1134 & 4.075 & 3.14 & ...    & 64 & 7.8 &4.4 &3.1 &2.6 &0.71 &0.51 &0.137&0.001\\
J1824$-$2452A & 3.054 & 119.86 & ... & 64 & 10.6&1.0 &2.0 &0.4 &0.33 &0.05 &0.972&0.003\\
J1857+0943   & 5.362 & 13.31 & 12.33 & 32 & 10.5&2.4 &5.0 &3.5 &1.01 &1.13 &0.518&0.002\\
J1909$-$3744 & 2.947 & 10.39 & 1.53  & 64 &6.1 &7.0 &2.1 &1.7 &0.77 &0.51 &0.0437&0.0002\\
J1939+2134   & 1.558 & 71.04 & ...   & 32 & 63 & 19 &13.2&5.0 &1.55 &0.72 &0.0382^a&0.0001^a\\
J2124$-$3358 & 4.931 & 4.62 & ...    & 32 & 12.3&11.8&3.6 &1.7 &0.44 &0.09 &0.524&0.006\\
J2129$-$5721 & 3.726 & 31.85 & 6.63  & 64 & 5.0 &2.1 &1.1 &0.7 &0.11 &0.05 &0.262&0.002\\
J2145$-$0750 & 16.052 & 9.00 & 6.84  & 64 &16.4 &8.9 &8.9 &12.5&1.38 &0.51 &0.337&0.002\\
\hline
$^a$APSR data
\end{tabular}}
\end{center}
\end{table*}

There are a number of on-going searches for pulsars at various
observatories around the world. When an MSP suitable for PTA
observations is discovered and its basic parameters measured with
sufficient accuracy, consideration is given to including it in the PTA
observations. From time to time, consideration is also given to
dropping one or more of the lesser-performing pulsars.  Specifically,
PSR J2241$-$5236 \citep{kjr+11} was added to the PPTA observing schedule
on 2010 February 9, PSR J1017$-$7156 \citep{kjb+12} was added on 2010
September 7 and PSR J1732$-$5049 was effectively dropped from the
schedule in 2011 January. Results for the two new pulsars are not
discussed in this paper. 

The PPTA sample of MSPs has been regularly observed with good quality
data at the three bands, 50cm, 20cm and 10cm, since early 2005. In
this paper we report on data obtained between 2005 March 1 (MJD 53430)
and 2011 February 28 (MJD 55620). Observation sessions are typically 2
-- 3 days in duration and at intervals of 2 -- 3 weeks. For PSRs
J1857+0943 and J1939+2134, shorter observation times were chosen
(Table~\ref{tb:ppta_psrs}) since these are northern sources which are
being monitored at other observatories. PSR J2124$-$3358 has a
timescale for diffractive scintillation at 1400 MHz which is longer
than the observation time and hence the pulsar is often not
visible. If there is sufficient observing time, having two shorter
observations separated by a time long compared to the scintillation
timescale increases the probability of obtaining a good
ToA. Generally, low-DM pulsars tend to be more affected by
scintillation since the scintillation bandwidth is comparable to or
larger than the observed bandwidth. Observers can terminate an
observation prematurely if there is a low probability of obtaining a
satisfactory ToA from the full observation time. This occurred
occasionally, mostly for 20cm observations as the scintillation
patterns are effectively uncorrelated for the 10cm and 50cm bands.

\subsection{Dispersion Corrections}\label{ssec:dmc}
For the PPTA pulsars, DM variations of up to 0.005 cm$^{-3}$~pc per
year have been observed \citep{yhc+07}. Since a DM variation of this
size introduces a variable time delay of order 10~$\mu$s in a 20cm ToA
it is clearly neccessary to correct for these variations.  Timing
analyses normally convert measured ToAs to infinite frequency using an
estimate of the DM. If one is concerned about DM variations, this DM
estimate may be obtained simultaneously with the timing analysis
provided multi-frequency data sets are available. To illustrate the
relevant issues we simplify the problem by assuming that the DM is
measured using ToAs $t_1$ and $t_2$ at two frequencies $f_1$ and
$f_2$, where $f_1$ is the primary frequency, i.e., the one with the
best quality ToAs. We use this DM to compute the infinite-frequency
ToA $t_{1,\infty}$ corresponding to $t_1$:
\begin{equation}
t_{1,\infty} = t_1(1+F) - t_2F
\end{equation}
where $F = 1/[(f_1/f_2)^2 -1]$. To ensure that the DM correction does
not add significant noise to $t_{1,\infty}$, we require that the uncertainty
in $t_2$, $\delta t_2$, be less than $[(1+F)/F]\delta t_1$. For the
PPTA observations, in the case where 20cm-band observations are
DM-corrected using 50cm-band data, $f_1/f_2 \approx 2.0$ and $F
\approx 1/3$, and so the 50cm ToAs need to have an
uncertainty less than four times that of the 20cm ToAs. For the
10cm/50cm combination, $f_1/f_2 \approx 4.4$, $F \approx 0.054$ and
the 50cm ToAs need to have an uncertainty less than 18 times that of
the 10cm ToAs. The benefits of a large $f_1/f_2$ ratio are
obvious. 

For the PPTA, the 50cm receiver is less sensitive than
either of the 20cm or 10cm systems; the ratios
of $B^{1/2}/S_{\rm sys}$ for 50cm:20cm:10cm are approximately (from
Table~\ref{tb:rcvr} where $B$ is the bandwidth) 1.0:3.3:4.8. The
relatively lower sensitivity of the 50cm system is largely compensated
for by the typically steep spectral index of pulsars. From
Table~\ref{tb:ppta_psrs}, the mean flux-density ratios across the 20
PPTA pulsars are $\langle S_{700}:S_{1400} \rangle \approx 3.0$ and
$\langle S_{700}:S_{3100} \rangle \approx 17.5$, corresponding to mean
spectral indices of $-1.57$ and $-1.93$ respectively. These ratios
suggest that the 50cm ToA uncertainties should not contribute
significantly to the uncertainty of the DM-corrected 20cm and 10cm
ToAs. However, other factors also need to be considered. RFI is
generally more of a problem at lower frequencies and so affects the
50cm data more than the other bands. Pulse widths are generally larger
at lower frequencies, either because of an intrinsic frequency
dependence of the pulse profiles or because of interstellar
scattering. Some pulsars have flatter than average spectra and so the
low-frequency ToAs have relatively greater uncertainties. Finally, ToA
variations not described by the dispersion relation
(Equation~\ref{eq:dm}) are sometimes observed at low frequencies. As a
consequence of these factors, correction for DM variations is not
always advantageous. These issues are discussed in more detail by
\citet{kcs+12}.

\section{Results and Discussion}\label{sec:results}

\subsection{Timing Data Sets}\label{sec:data}
ToAs produced by the processing pipeline described in
\S\ref{ssec:pipeline} are stored in the mySQL database and are
available for various timing analyses. As mentioned above, it is
common for two or more backend instruments to simultaneously process
the data from a given observation. ToAs based on different backend
systems processing the same input data are not independent and, for
each observation, only the ToA with the smallest uncertainty is
retained for subsequent analysis. For 50cm observations, the best
quality data are always obtained with the coherently dedispersing
instruments, initially CPSR2 and later APSR. For the higher-frequency
bands, the best data are normally obtained using the PDFB systems
which have wider bandwidths. Exceptions are the high DM/$P$ pulsars
J1824$-$2452A and J1939+2134 at 20cm, for which the best results are
usually obtained from APSR.

For some pulsars the MEM calibration made little difference to
the reduced $\chi^2$ of the timing solution whereas for others it
resulted in a large improvement. For example, for PSR J1744$-$1134,
the uncorrected 20cm rms timing residual is 0.50~$\mu$s and the
reduced $\chi^2$ is 11.0; for the corrected data the corresponding
numbers are 0.32~$\mu$s and 4.77. For simplicity, where the correction
made little difference, the uncorrected data were used.

Three of our pulsars have ecliptic latitudes less than $5\degr$: PSRs
J1022+1001 ($-0.06\degr$), J1730$-$2304 ($0.19\degr$) and J1824-2452A
($-1.55\degr$). ToAs from these pulsars are significantly delayed by
the solar wind when the path to them is close to the Sun. Standard
timing programs such as {\sc Tempo2} include a correction for this
dispersive delay. However, observations show that the actual delay
varies by a factor of two or more from one year to the next
\citep{yhc+07a,ychm12} and this variation is normally not
modelled. The effect of this on our results is discussed in
\S\ref{sec:best} below.

Figure~\ref{fg:3band} shows timing residuals for the final three-band
data sets for four of the PPTA pulsars. These illustrate the relative
timing precisions obtained in the different bands for different
pulsars, mainly depending on the pulsar spectral index and the effect
of DM variations on the timing. PSR J1045$-$4509 has large DM
variations \citep{yhc+07} and so there are significant residual
variations that approximately follow the $f^{-2}$ DM delay
dependence. For PSR J0613$-$0200 on the other hand, the DM-related
variations are small. It is also evident that there are
systematic offsets between the ToAs in the different bands for a given
pulsar. This is because the cross-correlation template alignment
procedure described in \S\ref{sec:offline} only approximates the
actual frequency dependence of the profile.

\begin{figure*}
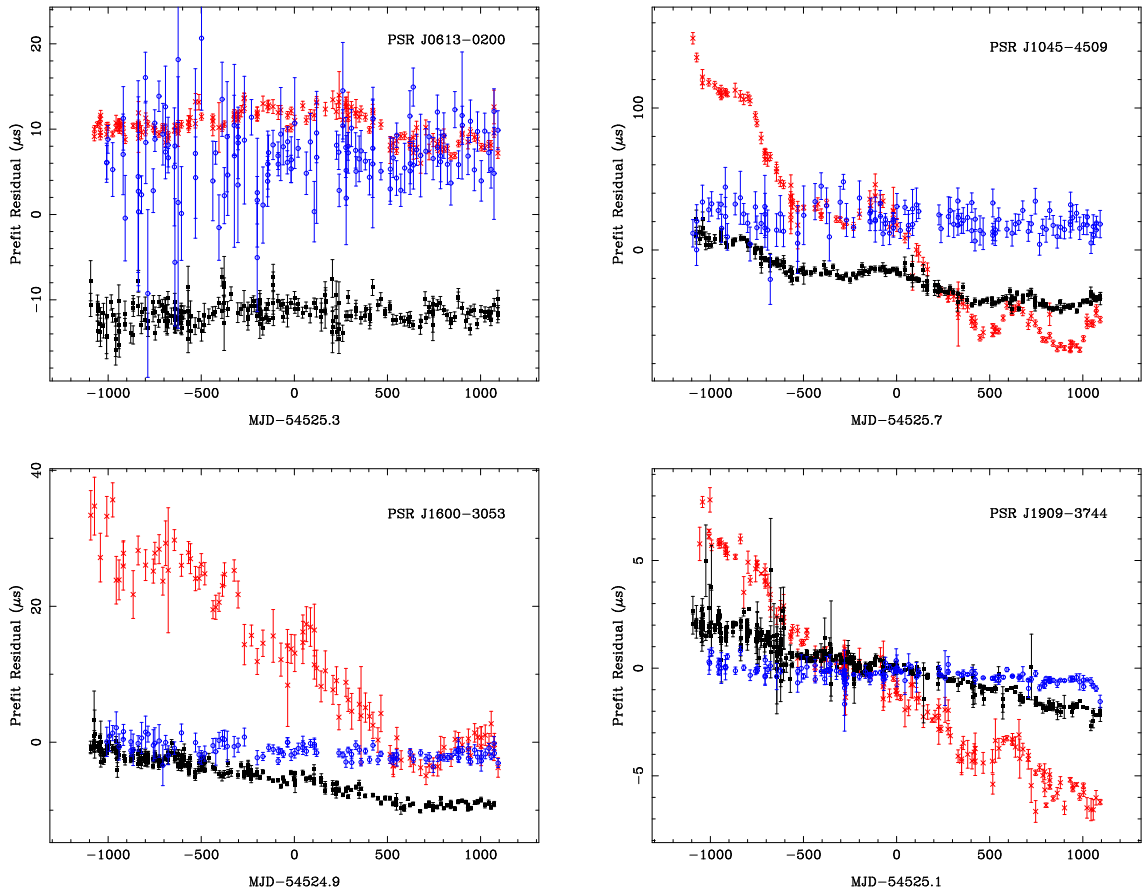

\begin{minipage}{150mm}
\begin{tabular}{cc}
\mbox{\psfig{file=J0613_dr1_res3.ps,angle=-90,width=75mm}} 
      & \mbox{\psfig{file=J1045_dr1_res3.ps,angle=-90,width=75mm}}\\
\mbox{\psfig{file=J1600_dr1_res3.ps,angle=-90,width=75mm}} 
      & \mbox{\psfig{file=J1909_dr1_res3.ps,angle=-90,width=75mm}}\\
\end{tabular}
\caption{Timing residuals for the three PPTA bands, 50cm (red
  $\times$), 20cm (black filled square) and 10cm (blue open circle)
  for four of the PPTA pulsars. Parameter files are from the 3-band
  solutions (see \S\ref{sec:dmv} below) with the DM corrections and
  interband jumps set to zero and all other parameters held fixed.}
\label{fg:3band}
\end{minipage}
\end{figure*}

Tables~\ref{tb:10cm_res} -- \ref{tb:50cm_res} summarise our timing
data sets for the three bands. In each table, after the pulsar name,
the next three columns are the full data span, the number of ToAs and
the median ToA uncertainty, respectively. To give an indication of our
current capabilities, the next four columns give results based on the
last year of the PPTA data sets, that is, from 2010 March to 2011
February inclusive. The columns are respectively, the number of ToAs
in the one-year data set, the weighted mean timing residual and the
corresponding reduced $\chi^2$, and the median ToA uncertainty for the
one-year set. Only the pulse frequency $\nu$ and its first two
derivatives, $\dot\nu$ and $\ddot\nu$ were fitted. The remaining
parameters, excepting the DM offsets which were set to zero, were held
at values obtained from the full three-band analysis described in
\S\ref{ssec:dmc} above. In a few cases, as noted in the tables, the
data spans were varied to give parameters which more closely
represented the system performance. For PSRs J1732$-$5049 and
J1824$-$2452A, the data spans were increased to get a sufficient
number of data points, and for PSR J1939+2134 they were decreased to
reduce the effect of the large DM variations on the rms residuals and
reduced $\chi^2$ values. 

\begin{table*}
\begin{center}
\caption{10cm-band timing results for the PPTA pulsars}\label{tb:10cm_res}
\begin{tabular}{lcd{0}d{5}cd{5}d{3}d{5}}
\hline 
\multicolumn{1}{c}{PSR} & Data span & \multicolumn{1}{c}{N$_{\rm ToA}$} 
&\multicolumn{1}{c}{Med. $\sigma_{\rm ToA}$}
& \multicolumn{1}{c}{N$_{\rm 1yr}$} & \multicolumn{1}{c}{1yr rms res.} 
&\multicolumn{1}{c}{$\chi^2_{\rm r}$}
&\multicolumn{1}{c}{Med. $\sigma_{\rm ToA,1yr}$} \\
& (MJD) & & \multicolumn{1}{c}{($\mu$s)} & & \multicolumn{1}{c}{($\mu$s)} 
& & \multicolumn{1}{c}{($\mu$s)} \\
\hline 
J0437$-$4715$^a$& 53880 -- 55619 & 475 &0.033 & 39 & 0.058& 4.77 & 0.027\\
J0613$-$0200  & 53517 -- 55619 & 130 & 2.67 & 23 & 2.61 & 1.44 & 2.37 \\
J0711$-$6830  & 53431 -- 55619 & 109 & 4.21 & 24 & 3.20 & 1.14 & 3.72 \\
J1022+1001    & 53431 -- 55619 & 127 & 1.60 & 22 & 3.44 &14.13 & 1.12 \\
J1024$-$0719  & 53431 -- 55619 & 109 & 5.00 & 20 & 2.76 & 0.68 & 4.62 \\
J1045$-$4509  & 53431 -- 55619 & 117 & 8.24 & 24 & 6.30 & 1.00 & 7.20 \\
J1600$-$3053  & 53517 -- 55619 & 104 & 0.78 & 24 & 0.58 & 1.01 & 0.69 \\
J1603$-$7202  & 53431 -- 55619 & 91  & 6.74 & 21 & 4.61 & 1.24 & 5.07 \\
J1643$-$1224  & 53431 -- 55619 & 98  & 1.64 & 18 & 1.51 & 1.59 & 1.35 \\
J1713+0747    & 53532 -- 55619 & 98  & 0.22 & 20 & 0.24 & 4.55 & 0.20 \\
J1730$-$2304  & 53451 -- 55599 & 83  & 3.11 & 16 & 0.60 & 0.74 & 2.21 \\
J1732$-$5049  & 53738 -- 55583 & 41  & 7.07 & 12$^b$ & 2.33 & 1.06 & 6.62 \\
J1744$-$1134  & 53467 -- 55620 & 84  & 0.85 & 20 & 0.37 & 1.17 & 0.66 \\
J1824$-$2452A & 53606 -- 55583 & 56  & 1.47 & 11$^c$ & 0.85 & 0.83 & 1.16 \\
J1857+0943    & 53452 -- 55620 & 65  & 2.70 & 16 & 1.16 & 1.06 & 2.79 \\
J1909$-$3744  & 53517 -- 55619 & 138 & 0.16 & 24 & 0.083& 1.47 & 0.13 \\
J1939+2134    & 53484 -- 55619 & 68  & 0.26 & 10$^d$ & 0.22 & 2.71 & 0.16 \\
J2124$-$3358  & 53482 -- 55619 & 95  & 8.05 & 21 & 6.48 & 0.95 & 7.31 \\
J2129$-$5721  & 53686 -- 55619 & 58  &17.70 & 19 & 14.6 & 0.90 & 22.1 \\
J2145$-$0750  & 53452 -- 55619 & 115 & 1.85 & 20 & 1.54 & 1.60 & 1.46 \\
\hline
\end{tabular}
a: Invariant interval. b: 2-year data span.  c: 1.5-year data span. d: 6-month data span.
\end{center}
\end{table*}

\begin{table*}
\begin{center}
\caption{20cm-band timing results for the PPTA pulsars}\label{tb:20cm_res}
\begin{tabular}{lcd{0}d{5}cd{5}d{3}d{5}}
\hline 
\multicolumn{1}{c}{PSR} & Data span & \multicolumn{1}{c}{N$_{\rm ToA}$} 
&\multicolumn{1}{c}{Med. $\sigma_{\rm ToA}$}
& \multicolumn{1}{c}{N$_{\rm 1yr}$} & \multicolumn{1}{c}{1yr rms res.} 
&\multicolumn{1}{c}{$\chi^2_{\rm r}$}
&\multicolumn{1}{c}{Med. $\sigma_{\rm ToA,1yr}$} \\
& (MJD) & & \multicolumn{1}{c}{($\mu$s)} & & \multicolumn{1}{c}{($\mu$s)} 
& & \multicolumn{1}{c}{($\mu$s)} \\
\hline 
J0437$-$4715$^a$& 53431 -- 55615 & 998 &0.038 & 95 & 0.087& 7.48 & 0.031 \\
J0613$-$0200    & 53431 -- 55620 & 218 & 0.88 & 31 & 0.81 & 1.71 & 0.63 \\
J0711$-$6830    & 53431 -- 55620 & 212 & 2.49 & 36 & 0.46 & 0.99 & 2.38 \\
J1022+1001$^b$  & 53468 -- 55618 & 246 & 1.17 & 26 & 0.66 & 3.78 & 0.84 \\
J1024$-$0719$^b$& 53431 -- 55620 & 175 & 1.74 & 28 & 1.06 & 1.48 & 2.10 \\
J1045$-$4509$^b$& 53450 -- 55620 & 185 & 2.13 & 31 & 1.75 & 1.04 & 1.81 \\
J1600$-$3053$^b$& 53430 -- 55598 & 237 & 0.50 & 22 & 0.44 & 2.76 & 0.29 \\
J1603$-$7202$^b$& 53430 -- 55619 & 168 & 1.00 & 27 & 1.10 & 2.42 & 0.83 \\
J1643$-$1224    & 53452 -- 55598 & 148 & 0.69 & 22 & 0.88 & 2.52 & 0.60 \\
J1713+0747      & 53452 -- 55598 & 198 & 0.16 & 23 & 0.12 & 5.86 & 0.076 \\
J1730$-$2304    & 53431 -- 55598 & 130 & 1.20 & 16 & 0.98 & 2.36 & 0.87 \\
J1732$-$5049    & 53724 -- 55582 & 102 & 2.08 & 19$^c$ & 1.47 & 0.78 & 1.85 \\
J1744$-$1134$^b$& 53452 -- 55599 & 169 & 0.38 & 20 & 0.17 & 1.71 & 0.30 \\
J1824$-$2452A   & 53518 -- 55620 & 178 & 0.48 & 15 & 0.57 & 5.65 & 0.32 \\
J1857+0943$^b$  & 53431 -- 55599 & 121 & 1.07 & 18 & 0.69 & 1.33 & 1.07 \\
J1909$-$3744    & 53431 -- 55620 & 396 & 0.13 & 33 & 0.097& 4.11 & 0.086 \\
J1939+2134      & 53450 -- 55599 & 139 & 0.12 & 9$^d$ & 0.096& 23.14& 0.035 \\
J2124$-$3358    & 53431 -- 55619 & 184 & 2.08 & 25 & 1.34 & 1.63 & 1.31 \\
J2129$-$5721$^b$& 53476 -- 55618 & 182 & 2.27 & 28 & 0.88 & 0.91 & 1.38 \\
J2145$-$0750$^b$& 53431 -- 55618 & 215 & 1.25 & 29 & 0.44 & 1.43 & 1.06 \\
\hline
\end{tabular}
a: Invariant interval. b: MEM calibration. c: 1.5-year data span. d: 6-month data span.
\end{center}
\end{table*}

\begin{table*}
\begin{center}
\caption{50cm-band timing results for the PPTA pulsars}\label{tb:50cm_res}
\begin{tabular}{lcd{0}d{5}cd{5}d{3}d{5}}
\hline 
\multicolumn{1}{c}{PSR} & Data span & \multicolumn{1}{c}{N$_{\rm ToA}$} 
&\multicolumn{1}{c}{Med. $\sigma_{\rm ToA}$}
& \multicolumn{1}{c}{N$_{\rm 1yr}$} & \multicolumn{1}{c}{1yr rms res.} 
&\multicolumn{1}{c}{$\chi^2_{\rm r}$}
&\multicolumn{1}{c}{Med. $\sigma_{\rm ToA,1yr}$} \\
& (MJD) & & \multicolumn{1}{c}{($\mu$s)} & & \multicolumn{1}{c}{($\mu$s)} 
& & \multicolumn{1}{c}{($\mu$s)} \\
\hline 
J0437$-$4715$^a$ &  53447 -- 55619 & 735 & 0.30 & 72 & 0.162& 2.83 & 0.14 \\
J0613$-$0200 &  53451 -- 55619 & 117 & 0.53 & 24 & 0.73 & 2.78 & 0.45 \\
J0711$-$6830 &  53431 -- 55619 & 127 & 4.15 & 26 & 2.11 & 1.38 & 3.12 \\
J1022+1001   &  53431 -- 55619 & 119 & 1.65 & 21 & 1.17 & 1.57 & 1.40 \\
J1024$-$0719 &  53431 -- 55619 & 75  & 4.00 & 20 & 1.07 & 0.58 & 2.83 \\
J1045$-$4509 &  53450 -- 55619 & 125 & 3.00 & 24 & 3.37 & 3.23 & 2.11 \\
J1600$-$3053 &  53431 -- 55619 & 97  & 1.84 & 20 & 1.21 & 0.90 & 1.43 \\
J1603$-$7202 &  53431 -- 55619 & 102 & 1.94 & 20 & 1.21 & 1.39 & 1.49 \\
J1643$-$1224 &  53431 -- 55619 & 91  & 1.47 & 19 & 1.89 & 3.58 & 1.16 \\
J1713+0747   &  53431 -- 55619 & 96  & 1.17 & 17 & 0.78 & 1.70 & 0.99 \\
J1730$-$2304 &  53451 -- 55599 & 84  & 2.03 & 16 & 1.43 & 1.70 & 1.46 \\
J1732$-$5049 &  53724 -- 55583 & 54  & 4.60 &  9$^b$ & 2.77 & 1.63 & 4.16 \\
J1744$-$1134 &  53432 -- 55620 & 82  & 0.66 & 20 & 0.65 & 4.59 & 0.37 \\
J1824$-$2452A&   53606 -- 55583 & 68  & 1.16 & 12$^c$ & 2.64& 11.04 & 1.16 \\
J1857+0943   &  53482 -- 55620 & 74  & 3.00 & 17 & 2.40 & 1.48 & 2.18 \\
J1909$-$3744 &  53468 -- 55619 & 156 & 0.30 & 24 & 0.39 & 7.10 & 0.24 \\
J1939+2134   &  53484 -- 55619 & 79  &0.060 & 10$^d$ & 0.42 & 42.15& 0.089 \\
J2124$-$3358 &  53452 -- 55619 & 84  & 3.64 & 21 & 1.27 & 0.97 & 2.33 \\
J2129$-$5721 &  53686 -- 55619 & 105 & 1.32 & 22 & 1.31 & 2.60 & 1.09 \\
J2145$-$0750 &  53468 -- 55619 & 114 & 1.48 & 20 & 0.71 & 0.63 & 1.31 \\
\hline
\end{tabular}
a: Invariant interval. b: 2-year data span.  c: 1.5-year data span. d: 6-month data span.
\end{center}
\end{table*}

ToA uncertainties are as computed by the template fitting program, in
our case the {\sc Psrchive} routine {\sc pat} using the \citet{tay90a}
algorithm. No scaling or biasing (``EFAC'' or ``EQUAD'') factors have
been applied. The tables show that in most cases the median ToA
uncertainties are greater for the full six-year data sets than they
are for the one-year data sets. This is especially true for pulsars
with high DM/$P$ and mostly results from the improvement in back-end
systems over the six years.

It is notable that many of the reduced $\chi^2$ values for the
one-year data sets are close to 1.0, especially for the 10cm data sets
and for the weaker pulsars. This indicates that the computed ToA
uncertainties are generally accurate. In some cases, the reduced
$\chi^2$ values are less than 1.0. This primarily occurs in pulsars
and bands that have large intensity fluctuations due to interstellar
scintillation and hence a wide range of weights in the least-squares
fit. This effectively reduces the number of degrees of freedom for the
fit and can result in an under-estimate of the corresponding rms
residual. However, the minimum reduced $\chi^2$ value (0.58 for PSR
J1024$-$0719 at 50cm) is still reasonably close to 1.0 and so the
effect is not very significant.

For the stronger pulsars, at 20cm and 50cm especially, the reduced
$\chi^2$ values tend to be larger than 1.0, indicating short-term
timing noise in excess of that expected from the ToA
uncertainty. Possible reasons for this include residual RFI,
especially at the lower frequencies, short-term variations in
interstellar dispersion or scattering and short-term variations in the
intrinsic profile shape (commonly known as ``pulse jitter'').  In some
cases, e.g., PSRs J0437$-$4715 and J1939+2134, the median ToA
uncertainties are very small, $\lapp 30$ns, and so reveal
perturbations that are not obvious in the weaker pulsars. For PSR
J0437$-$4715 it is likely that most of additional scatter results from
pulse jitter \citep{ovh+11}.  Errors in the formation of the invariant
interval used for timing this pulsar are also possible. Although the
invariant interval is nominally unaffected by calibration errors
\citep{bri00}, various second-order effects may be important. The
invariant interval is essentially the difference between Stokes $I$
and the polarised component $P$. $P$ is a positive-definite quantity
and suffers a noise bias in the same way as the linearly polarized
component $L=(Q^2+U^2)^{1/2}$ \citep{lk05}. This means that the
invariant-interval profile is dependent on the S/N of the polarisation
profiles used to form it. To minimise this effect, we summed the PSR
J0437$-$4715 profiles to eight sub-integrations and 32 frequency
channels before forming the invariant intervals. Because of this
issue, we did not use invariant-interval profiles for any of the other
PPTA pulsars.

PSR J1022+1001 is an interesting case. \citet{kxc+99} found variations
in the relative amplitudes of profile pulse components on timescales
of order hours and showed that such changes would induce offsets in
derived ToAs. Their timing solutions had rms residuals of 20 -- 100
$\mu$s, substantially more than expected from random baseline noise,
and they attributed the excess noise to the profile
variations. However, \citet{hbo04} used carefully calibrated CPSR2
data recorded at Parkes and obtained an rms timing residual of just
2.27 $\mu$s in 5-minute integrations. They found no evidence for
significant variations in profile shape on this or longer
timescales. The 1-year 20cm MEM-calibrated PPTA data set has an
rms timing residual of 0.66 $\mu$s (Table~\ref{tb:20cm_res}). This
shows that the effect of any profile variations is very small, much
less than those observed by \citet{kxc+99}. Never-the-less, the reduced
$\chi^2$ of 3.78 indicates that systematic ToA offsets are still
present and that these have a timescale of the order of hours. As
Figure~\ref{fg:tmplt} shows, PSR J1022+1001 has a huge
frequency-dependent variation of profile shape. It also scintillates
strongly, especially at 20cm (Table~\ref{tb:ppta_psrs}). The pulse
signal is therefore likely to be strong in different parts of the band
at different times, resulting in a variable bias to the measured
ToAs. This may dominate the ToA scatter that we observe. It
is also possible that intrinsic temporal profile shape variations play a role,
but a much smaller one than suggested by \citet{kxc+99}. The pulse
profile of PSR J1022+1001 has very high linear polarisation, nearly
100\% in the trailing half \citep{kxc+99,ymv+11}. It seems
most probable that the large profile and timing variations observed by
\citet{kxc+99} resulted from errors in polarisation calibration.

Strong scintillation coupled with a frequency-dependent pulse profile
may play a significant role for several of our pulsars. This effect
could be largely eliminated by use of frequency-dependent templates,
but it requires a careful study to assess its importance to the
short-term timing noise relative to other contributions. For example,
for PSRs J1824$-$2452A and J1939+2134 there are large DM variations,
the effects of which are not absorbed by the
$\nu$--$\dot\nu$--$\ddot\nu$ fitting. Residual RFI also may make a
significant contribution to the observed ToA scatter, especially at
50cm. Finally, instrumental errors and deficiencies in
signal-processing methods may contribute. These poorly understood
contributions to our timing residuals are the subject of continuing
investigation. With an improved understanding, we should be able to
reduce their effect and hence obtain more accurate data sets.

\subsection{Flux Densities and Pulse Widths}\label{sec:flux}
The mean and rms flux densities given in Table~\ref{tb:ppta_psrs} were
computed by taking PDFB observations (which have a more reliable
flux-density calibration) corresponding to ToAs from 2010 January to
the end of the data set. The flux densities were averaged in time over
each observation and in frequency over each band. For PSR
J0437$-$4715, the total-intensity (Stokes $I$) profile was used; the
invariant-interval mean flux densities are approximately 60\%, 85\%
and 80\% of the Stokes $I$ values at 10cm, 20cm and 50cm,
respectively. The fractional rms variations (or modulation indices) of
the measured flux densities for each band are often large, especially
for the low-DM pulsars. As mentioned in \S\ref{sec:obs}, for such
pulsars, observations when the pulsar had low flux density were
sometimes terminated early, especially at 20cm. The 20cm flux
densities quoted in Table~\ref{tb:ppta_psrs} are consequently somewhat
biased to high values, but this effect is small, typically a few
percent or less.

Table~\ref{tb:ppta_psrs} also gives mean pulse widths at 50\% of the
profile peak and the estimated uncertainty in these widths. These
results are averages of widths measured from the observation-averaged
20cm total-intensity profiles. For the PSR J0437$-$4715
invariant-interval profiles, the mean width is $0.1370\pm 0.0004$~ms,
approximately 3\% smaller than the Stokes $I$ width.

\subsection{Dispersion Variations}\label{sec:dmv}

DM variations have been determined for all 20 PPTA pulsars using the
three-band data as described in detail by \citet{kcs+12}. Here we give
a brief summary of the process. {\sc Tempo2} allows fitting to
multiband data sets for a set of offsets from the nominal DM at
specified times. The offsets are constrained to have zero mean and the
average DM is obtained from the same fit. The sample interval was
chosen to minimise the noise contributed to the DM-corrected residuals
by estimation errors in the DM offsets. The optimal interval for each
pulsar was taken to be the inverse of the ``corner'' frequency at
which the power in the DM variations equals the white noise level in
the ``best'' data set (see below). To obtain DM offsets at other
times, {\sc Tempo2} interpolates between the sample values. Simply
fitting for the DM offsets in this way will absorb some of the
frequency-independent variations that are crucial to PTA
objectives. To overcome this problem, we chose to additionally fit for a
``common-mode'' frequency-independent offset sampled at the same
intervals as the DM offsets. This common-mode signal can be used for
PTA purposes, but for this paper, we only use it to provide an
unbiased estimate of the DM variations. 

\begin{figure*}
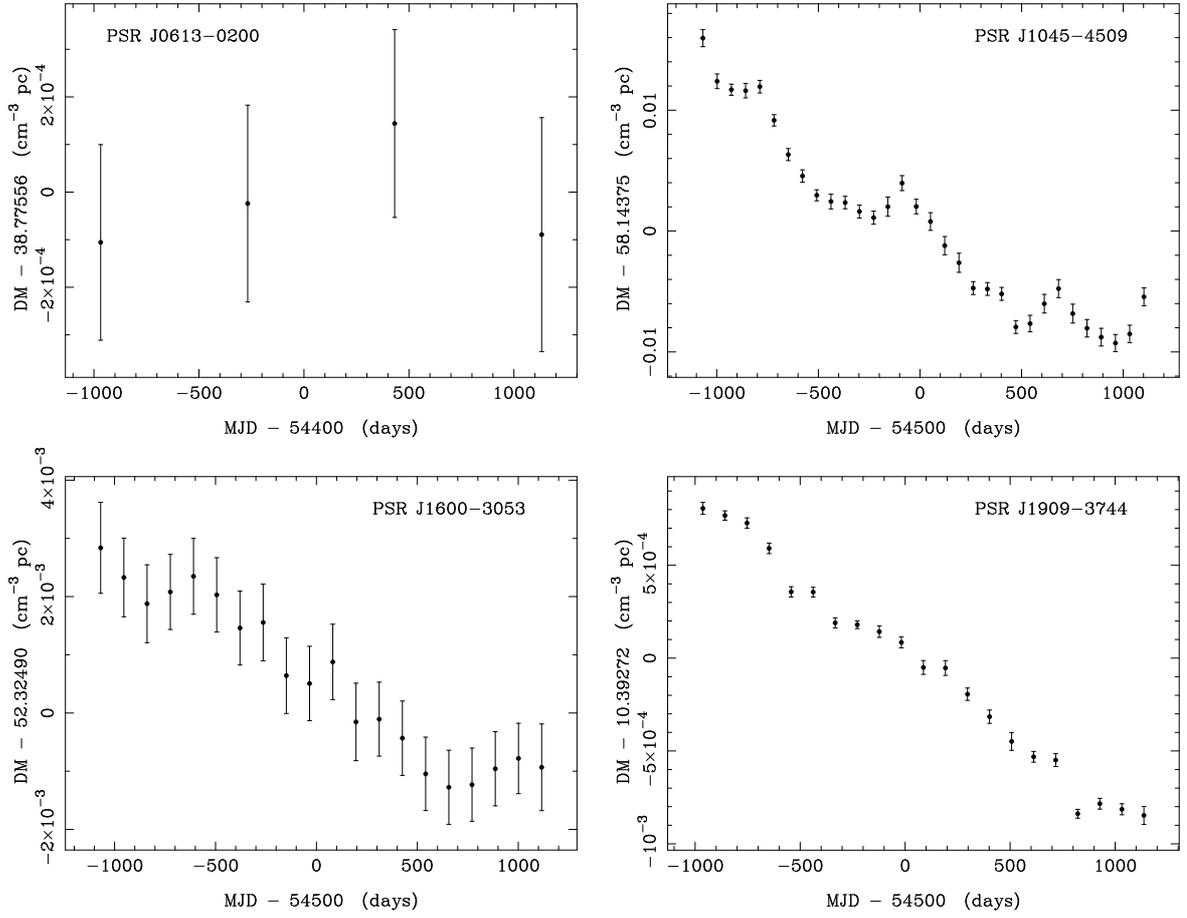

\begin{minipage}{150mm}
\begin{tabular}{cc}
\mbox{\psfig{file=J0613_dr1_dmc.ps,angle=-90,width=75mm}} 
      & \mbox{\psfig{file=J1045_dr1_dmc.ps,angle=-90,width=75mm}}\\
\mbox{\psfig{file=J1600_dr1_dmc.ps,angle=-90,width=75mm}} 
      & \mbox{\psfig{file=J1909_dr1_dmc.ps,angle=-90,width=75mm}}\\
\end{tabular}
\caption{Observed dispersion-measure (DM) variations for four PPTA pulsars}
\label{fg:dmvar}
\end{minipage}
\end{figure*}

Because it is impossible to align template profiles for the different
bands with sufficient accuracy, it is also necessary to simultaneously
fit for offsets between bands (with one chosen as reference). A single
offset per band pair is fitted. In the fit, ToAs are weighted by the
inverse square of their uncertainty. Inclusion of other free
parameters in the timing model depends on ToA precisions and the
physical properties of the pulsar, such as its distance and the nature
of the binary system. Cholesky whitening \citep{chc+11} is used to
properly account for the correlation in the residuals and thus to
provide reliable error estimates for the DM variations and other
parameters.  Figure~\ref{fg:dmvar} shows the DM variations derived
from the ToAs shown in Figure~\ref{fg:3band} using the above
procedure.

\subsection{Single-band Corrected Timing Data Sets}\label{sec:best}

\begin{figure*}[ht]
\centerline{\psfig{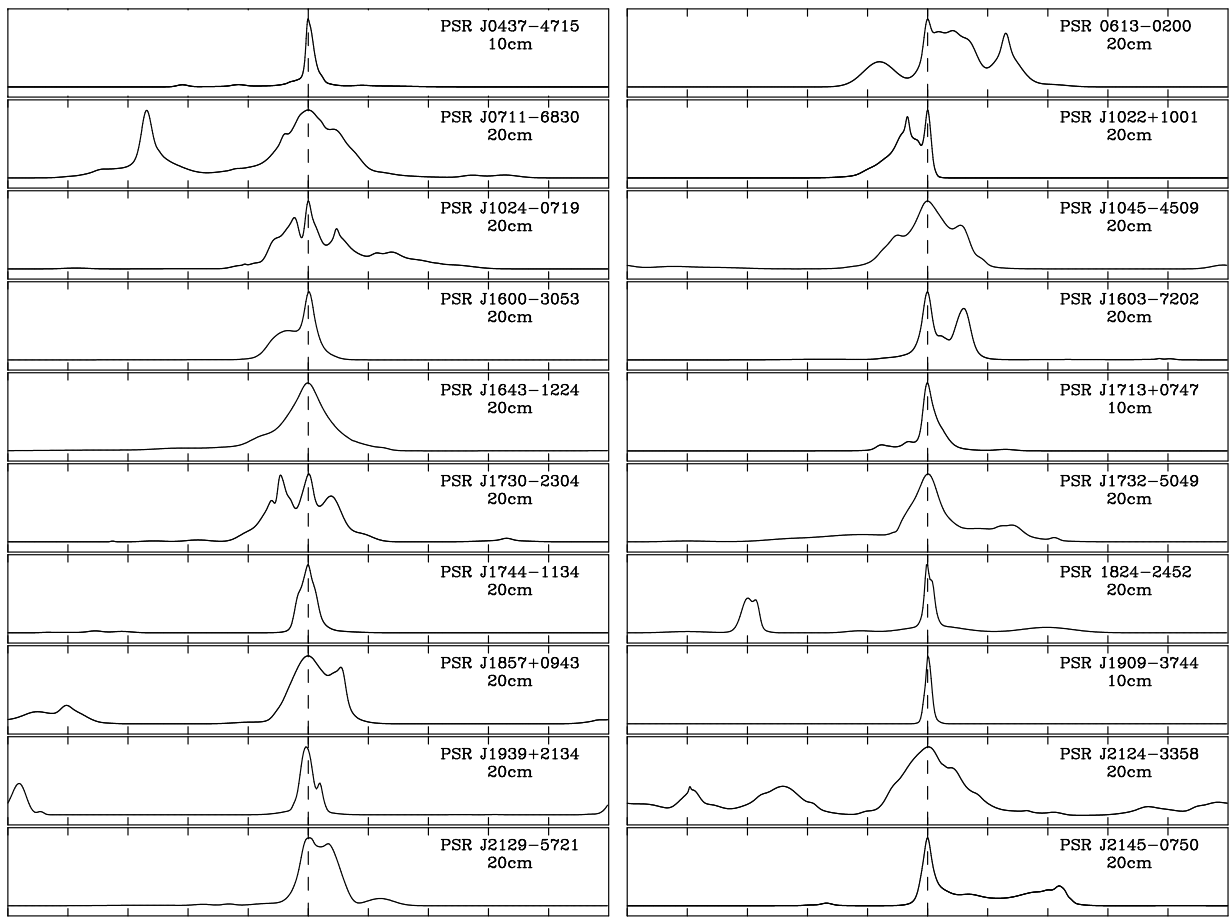}}
\caption{Analytic (noise-free) timing profile templates for the
  ``best'' band for each of the 20 PPTA pulsars. The full pulse period
  is shown in each case and the vertical dashed
  line gives the template reference phase. }\label{fg:tmplt_best}
\end{figure*}

\begin{figure*}[h]
\centerline{\psfig{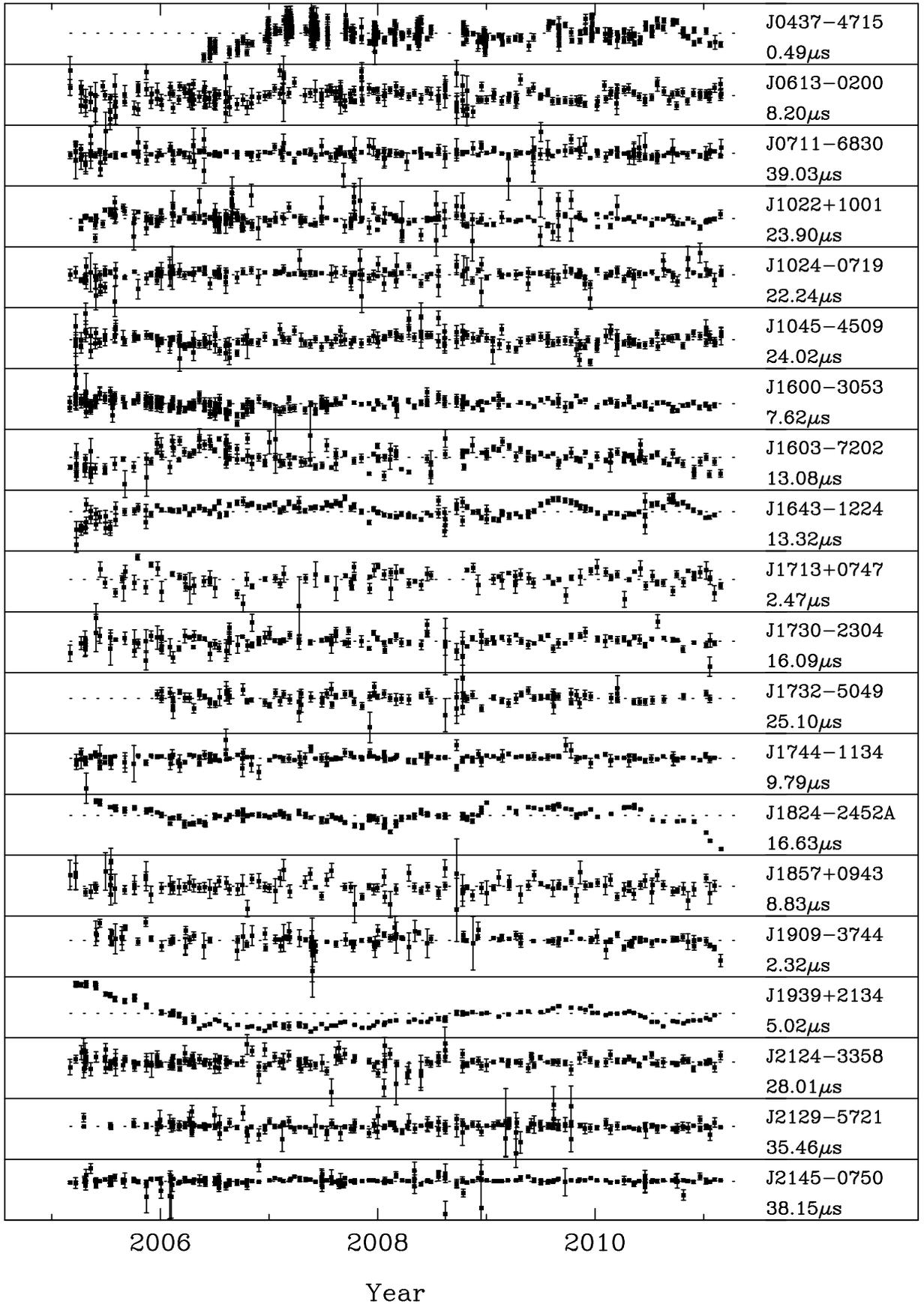}}
\caption{Final post-fit timing residuals for the PPTA pulsars for the
  band and corrections as listed in Table~\ref{tb:best}. The
  vertical extent of each subplot is adjusted to fit the data and its
  value is given below the pulsar name. The dashed line marks zero residual. }\label{fg:dr1_res}
\end{figure*}

As a basis for applications of the PPTA data set, ToAs for the band
having the lowest over-all rms timing residual, either with or without
the MEM calibration and with or without correction for DM variations,
were selected. Table~\ref{tb:best} gives this ``best'' band, the
correction procedure adopted and the data span in years for each PPTA
pulsar. Figure~\ref{fg:tmplt_best} shows the profile templates used
for the best instrument of the best band. The best instrument was
PDFB4 in all cases except two. For PSRs J1824$-$2452A and J1939+2134
APSR was used because their high DM/$P$ requires the coherent
dedispersion provided by this instrument. The template reference phase
is shown on each plot; knowledge of this allows comparison of the
absolute pulse ToAs from the PPTA with data from other telescopes.
For the 3-band Cholesky fit described in \S\ref{sec:dmv} the pulsar
parameters, including pulse frequency $\nu$ and $\dot\nu$, interband
jumps and DM corrections were simultaneously fitted. The fifth column
of Table~\ref{tb:best} gives the number of pulsar parameters fitted
and the sixth column gives, where applicable, the averaging interval
for the DM corrections. The seventh and eighth columns give the rms
timing residual and reduced $\chi^2$ after a fit of just $\nu$ and
$\dot\nu$ to the best-band data, with all other parameters, including
DM offsets where applied, held fixed at the values resulting from the
3-band fit.  Since the best-band data sets have just a single band,
there are of course no interband jumps. Figure~\ref{fg:dr1_res} shows
the post-fit timing residuals for these best-band fits.

\begin{table*}[h]
\begin{center}
\caption{Timing Results for the PPTA pulsars}\label{tb:best}
\begin{tabular}{lcccccd{5}d{3}}
\hline 
\multicolumn{1}{c}{PSR} & Band & Corr. & Data span & \multicolumn{1}{c}{N$_{\rm par}$} & 
DMC Int. &\multicolumn{1}{c}{Rms Res.} 
&\multicolumn{1}{c}{$\chi^2_{\rm r}$}  \\
& & & (yr) & & \multicolumn{1}{c}{(d)} &\multicolumn{1}{c}{($\mu$s)} & \\
\hline 
J0437$-$4715 & 10cm & {\sc ivi+dmc} & 4.76 &18 & 60 & 0.075& 5.50 \\
J0613$-$0200 & 20cm & {\sc dmc}     & 6.00 &13 &700 & 1.07 & 1.76   \\
J0711$-$6830 & 20cm & ...           & 6.00 &7  &... & 0.89 & 1.66   \\
J1022+1001   & 20cm & {\sc mem}     & 5.89 &12 &... & 1.72 & 9.27   \\
J1024$-$0719 & 20cm & {\sc mem}     & 6.00 &8  &... & 1.13 & 1.40   \\
J1045$-$4509 & 20cm & {\sc mem+dmc} & 5.94 &13 & 70 & 2.77 & 1.80   \\
J1600$-$3053 & 20cm & {\sc mem+dmc} & 5.94 &13 &115 & 0.68 & 2.78  \\
J1603$-$7202 & 20cm & {\sc mem}     & 6.00 &12 &... & 2.14 & 7.93   \\
J1643$-$1224 & 20cm & ...           & 5.88 &14 &... & 1.64 & 5.46   \\
J1713+0747   & 10cm & ...           & 5.71 &16 &... & 0.31 & 4.00  \\
J1730$-$2304 & 20cm & {\sc dmc}     & 5.94 &7  &300 & 1.47 & 2.90   \\
J1732$-$5049 & 20cm & {\sc dmc}     & 5.09 &12 &600 & 2.22 & 1.34  \\
J1744$-$1134 & 20cm & {\sc mem+dmc} & 5.88 &8  &1000& 0.32 & 4.77  \\
J1824$-$2452A& 20cm & {\sc dmc}     & 5.76 &7  &82.5& 2.44 & 30.22   \\
J1857+0943   & 20cm & {\sc mem+dmc} & 5.94 &12 &425 & 0.84 & 1.16   \\
J1909$-$3744 & 10cm & {\sc dmc}     & 5.76 &17 &105 & 0.133 & 2.21   \\
J1939+2134   & 20cm & {\sc dmc}     & 5.88 &7  & 50 & 0.68 & 141.63   \\
J2124$-$3358 & 20cm & {\sc dmc}     & 6.00 &8  &1800& 1.90 & 1.38   \\
J2129$-$5721 & 20cm & {\sc mem+dmc} & 5.87 &12 &1500& 0.80 & 1.00   \\
J2145$-$0750 & 20cm & {\sc mem}     & 6.00 &14 &... & 0.78 & 3.18   \\
\hline
\end{tabular}
\end{center}
\end{table*}

For three of the 20 pulsars we obtain our best timing by using
10cm-band data. There are two main reasons why higher frequencies will
give more precise ToAs. The first is that profile components generally
tend to be narrower at higher frequencies, leading to improved timing
precision. The second and usually more important reason is that
interstellar propagation effects are less significant at higher
frequencies. Dispersion delays vary as $f^{-2}$ and scintillation
effects more like $f^{-4}$. We attempt to correct for DM variations, but
these corrections are always imprecise and the error induced in the
residuals follows the same $f^{-2}$ dependence. 

Scintillation effects, especially those due to refractive
scintillation, are much more difficult to measure and correct for
\citep{hs08,crg+10,dem11}. This is especially true for MSPs and up to
now such techniques have not been successfully applied to this class
of pulsar. Scattering delays have long-term variations as the path to
the pulsar moves through the interstellar medium because of the
relative transverse motion of the scattering medium, the pulsar and
the Earth. However, diffractive scintillation also introduces
short-term variations. Mean pulse profiles for most pulsars are
significantly frequency dependent. Consequently, changes in the
relative pulsar flux density in different parts of the band will
directly result in variable ToA offsets. This will vary on the
timescale for diffractive scintillation. As indicated by the large rms
variations in observed flux density given in Table~\ref{tb:ppta_psrs},
this is either comparable to or larger than the observation time for
many of the PPTA pulsars and bands. Refractive scintillation also
introduces a variable delay due to variations in the effective path
length as different parts of the scattering screen are focussed on the
Earth. This is a broad-band effect and is not accompanied by any
significant change in profile shape. It is therefore very difficult to
predict and hence to correct for.
  
The pulsars for which we obtain our best timing at 10cm are among the
strongest PPTA pulsars for this band (Table~\ref{tb:ppta_psrs}). This
illustrates the fact that the primary reason that the 20cm band is
superior to the 10cm band for most of our pulsars, despite the factors
discussed above, is simply S/N. Because of the relatively steep
spectrum of most pulsars, the low S/N obtained at 10cm outweighs the
benefits of the higher frequency. This is also illustrated by the
large rms timing residuals we obtain at 10cm (Table~\ref{tb:10cm_res})
for many of our pulsars. Because of these factors, it is likely that
the best PTA-related use of more sensitive telescopes will be to allow
more pulsars to be observed at higher radio frequencies.

Nine of the 20 pulsars benefitted from the MEM calibration to the 20cm
data and 14 of the 20 benefitted from correction for DM
variations. For four pulsars neither correction gave a significant
improvement in the rms residuals. In principle, the MEM calibration
should always improve the stability of the profiles since it removes
the parallactic angle dependence of the calibrated profiles
\citep{van04c}. In practice, the correction made little difference to
the more weakly polarised pulsars and in these cases it was not
applied. For PSR J0437$-$4715, results with the MEM calibration were
inferior to those obtained by using the invariant interval. This
pulsar is a special case in that the polarisation flips between
orthogonal states right at the peak of the profile \citep[see,
e.g.,][]{ymv+11}. The ToAs for this pulsar are therefore extremely
sensitive to any errors in the calibration which alter the intensity
ratio of the orthogonal states at the pulse peak. The invariant
interval is relatively immune to calibration errors and in this case
gives the best results.

In most cases, rms residuals were reduced by applying the correction
for DM variations. For PSRs J1603$-$7202 and J1643$-$1224, systematic
frequency-dependent residual variations that did not follow the
$f^{-2}$ DM law were observed. For both pulsars, applying the derived
DM correction made the 20cm rms post-fit residuals significantly
worse. In both cases, the intervals of non-DM residual variation were
isolated and correlated in time over about six months, in 2006 for
J1603$-$7202 and in 2010 for J1643$-$1224. It is possible that these
were episodes of significant refractive delay but further investigation
is needed. It is interesting to note that an extreme scattering event
lasting three years and centred in 1998 was identified in flux-density
data for PSR J1643$-$1224 by \citet{mlc03}.

As mentioned in \S\ref{sec:data} above, solar-wind delays may be
significant for at least three of the PPTA pulsars. Unmodelled delays are
typically a few microseconds or less. However, since we do not always
observe when a pulsar is close to the Sun, they affect just 5 -- 10
ToAs even in the pulsars that pass within $5\degr$ of the Sun. We have
chosen to include these ToAs in our data sets to allow further
investigation of the solar-wind effects. The effect on the timing
results is small. For the most-affected pulsar, PSR J1022+1001,
eliminating ToAs for times when the pulsar was within $5\degr$ of the Sun
from the single-band data set (Table~\ref{tb:best}) reduces the rms
residual from 1.72~$\mu$s to 1.64~$\mu$s and the reduced $\chi^2$ from
9.27 to 8.85. For all other pulsars, the effect is less significant.

\subsection{Long-term pulse frequency variations}\label{sec:f2}
The results presented in Figure~\ref{fg:dr1_res} and comparison of the
6-year and 1-year rms timing residuals in Tables~\ref{tb:10cm_res} --
\ref{tb:best} show that, for a 6-year data span, about half of the
PPTA pulsars have significant long-term fluctuations in their pulse
frequency. These fluctuations may be quantified in different ways. A
simple approach is to fit a pulse frequency second derivative
($\ddot\nu$) to each data set; this is the essence of the timing
stability parameter $\Delta(t)$ \citep{antt94} and quantifies the
presence of a cubic term in the residuals. In Table~\ref{tb:f2} we
present $\ddot\nu$ values for the PPTA pulsars obtained in two
different ways. The second column for each pulsar lists the $\ddot\nu$
value obtained from a simultaneous fit of all parameters as described
in \S\ref{sec:dmv}, together with $\ddot\nu$, to the 3-band data
sets. In the third column for each pulsar we list the $\ddot\nu$ value
obtained by adding it to the best-band fit as described in
\S\ref{sec:best}, that is, a simultaneous fit of $\nu$, $\dot\nu$ and
$\ddot\nu$ to the best-band data set for each pulsar. 

In general, the $\ddot\nu$ values from the 3-band fits have a lower
significance than the values obtained by fitting to the best-band data
alone. There are several reasons for this. Firstly, we are fitting for
more parameters and some of these have significant covariances which
are reflected in the derived uncertainties. Secondly, even though we
try to minimise the effect by averaging over suitably chosen
intervals, fitting and correcting for DM variations always adds
effectively white noise to radio-frequency ToAs because the DM
corrections are always uncertain at some level. Thirdly, our procedure
for fitting $\ddot\nu$ to the best-band data may under-estimate the
true uncertainty in this parameter where DM corrections are applied
since some correlation is introduced into the effective ToA errors
because of the averaging of the DM corrections. This is not an issue
for the seven pulsars where DM corrections are not applied to the
best-band data sets (Table~\ref{tb:best}). In these cases the fitting
procedure is absolutely standard and parameters and their
uncertainties should be reliable.

In almost all cases, the $\ddot\nu$ values derived by the two methods
are in agreement within the combined uncertainties. A notable
exception is PSR~J0437$-$4715. Because of its large flux density
(Table~\ref{tb:ppta_psrs}) very precise ToAs are in principle
measurable for this pulsar. However, it is also true that observations
of it reveal systematic errors most clearly. Unfortunately, as
Figure~\ref{fg:dr1_res} shows, at its best band (10cm), data prior to
mid-2006 were not of the same quality as subsequent data and therefore
were not included in the data set. Earlier data at 20cm and 50cm are
included in the 3-band data set and so the two data sets are not
directly comparable, leading to the significantly different $\ddot\nu$
values in Table~\ref{tb:f2}.

Half of the PPTA pulsars have a $\ddot\nu$ value that is significant
at the $3\sigma$ level based on the best-band fits. While there are
some caveats about the uncertainty estimations as discussed above, we
believe the overall result is solid. As an illustration, we take the
important case of PSR~J1909$-$3744, where we find a significant
$\ddot\nu$ in contrast to other published results
\citep{vbc+09,dfg+12} where no red noise was detected in data sets of
similar duration.\footnote{In fact, in his thesis \citet{ver09}, did
  quote a $\ddot\nu$ for PSR J1909$-$3744,
  $(11\pm4)\times10^{-28}$~s$^{-3}$, which is consistent with the
  value in Table~\ref{tb:f2}} From Table~\ref{tb:f2}, $\ddot\nu$
values for the 3-band fit and the best-band fit are respectively
$7.8\pm3.5$ and $10.1\pm2.1 \times 10^{-28}$~s$^{-3}$. If we fit all
17 pulsar parameters plus $\ddot\nu$ to the best-band data including
the DM corrections held fixed, we obtain $\ddot\nu = 14.6\pm2.5 \times
10^{-28}$~s$^{-3}$. If we omit the DM corrections, giving a standard
multi-parameter fit to single-band data, we obtain $\ddot\nu =
14.5\pm2.6 \times 10^{-28}$~s$^{-3}$. All of these values are
consistent to $1.6\sigma$ or better and all the best-band fits give a
value of $\ddot\nu$ that is significant at about $5\sigma$.

These results represent the first detections of red timing
noise in a significant sample of MSPs. Red timing noise has long been
recognised in PSR J1939+2134 \citep[e.g.,][]{ktr94} and PSR
J1824$-$2452 \citep{hlk+04} -- these pulsars are anomalous in having
much stronger red noise than other MSPs. Indications of long-term
correlated residual fluctuations have been seen in MSPs J1713+0747
\citep{sns+05} and J0613$-$0200, J1024$-$0719, J1045$-$4509
\citep{vbc+09}, but no $\ddot\nu$ values were derived.

\citet{sc10} chose to quantify the noise properties of pulsars using
the rms timing residuals rather than $\ddot\nu$. They modeled the rms
timing residuals as a function of $\nu$, $\dot\nu$ and data span for
different samples of pulsars. We have compared the rms timing
residuals in Table~\ref{tb:best} with the values predicted by their
``CP+MSP'' model, which is based on a fit (in log space) to published
rms timing residuals for both normal (non-millisecond) pulsars and
MSPs.  They define a parameter $\zeta$ which is the observed rms
timing residual after fitting for $\nu$ and $\dot\nu$ divided by the
predicted value. For the PPTA pulsars, most $\zeta$ values are of
order ten or greater, indicating that the observed rms residual is
dominated by white noise. Exceptions are PSRs J0437$-$4715,
J1824$-$2452A, J1909$-$3744 and J1939+2134, for which the
$\zeta$-values are 0.51, 0.17, 1.68 and 0.54, respectively. Except for
PSR J1824$-$2452A, the observed rms residual is within a factor of two
of the prediction, which is satisfactory. There is some evidence that
the model over-estimates the timing noise in PSR J1824$-$2452A. Based
on the spread of timing noise measured by \citet{sc10}, the value of
$\zeta$ for this pulsar is a $1\sigma$ outlier. The model prediction
for PSR J1824$-$2452A is high largely because of the large
$|\dot\nu|$.  This pulsar is unique in the PPTA sample
in that it is associated with a globular cluster and furthermore is
within the core of the cluster \citep{lbm+87}, suggesting that the
large $|\dot\nu|$ may partly result from acceleration of the pulsar in
the cluster gravitational field. Consequently, no conclusion can yet
be drawn about whether or not its timing noise is anomalously low.  

\begin{table*}

\begin{center}
\caption{Pulse frequency second-time-derivatives for PPTA pulsars}\label{tb:f2}
\begin{tabular}{lcclcc}
\hline 
\multicolumn{1}{c}{PSR} & $\ddot\nu$ (3-band) & $\ddot\nu$ (best) & \multicolumn{1}{c}{PSR} & $\ddot\nu$ (3-band) & $\ddot\nu$ (best) \\
& ($10^{-28}$ s$^{-3}$) & ($10^{-28}$ s$^{-3}$) & & ($10^{-28}$ s$^{-3}$) & ($10^{-28}$ s$^{-3}$) \\
\hline 
J0437$-$4715 &$-1.9\pm0.9$&$-5.3\pm0.6$& J1730$-$2304 & $-11\pm10$ & $-2\pm9$   \\
J0613$-$0200 & $1\pm14$   & $8\pm12$   & J1732$-$5049 & $55\pm57$  & $6\pm31$   \\
J0711$-$6830 & $-12\pm8$  & $-23\pm6$  & J1744$-$1134 &$5.7\pm3.8$ &$7.9\pm3.3$ \\
J1022+1001   &$10.6\pm2.9$& $9.4\pm3.1$& J1824$-$2452A& $392\pm48 $& $314\pm23$ \\
J1024$-$0719 & $-40\pm8$  & $-35\pm7$  & J1857+0943   & $-2\pm14$  & $4\pm8$    \\
J1045$-$4509 & $6 \pm44$  & $20\pm14$  & J1909$-$3744 &$7.8\pm3.5$ &$10.1\pm2.1$\\
J1600$-$3053 & $30\pm17$  & $39\pm6$   & J1939+2134   & $231\pm22$ & $216\pm11$ \\
J1603$-$7202 & $-22\pm7$  & $-3\pm6$   & J2124$-$3358 & $-13\pm15$ & $5\pm14$   \\
J1643$-$1224 &$-100\pm37$ & $-53\pm15$ & J2129$-$5721 & $26\pm14$  & $13\pm9$   \\
J1713+0747   & $12\pm4$   & $8\pm4$    & J2145$-$0750 &$11.9\pm3.3$& $7.4\pm1.7$\\
\hline
\end{tabular} 
\end{center} 
\end{table*}

\begin{figure*}
\centerline{\psfig{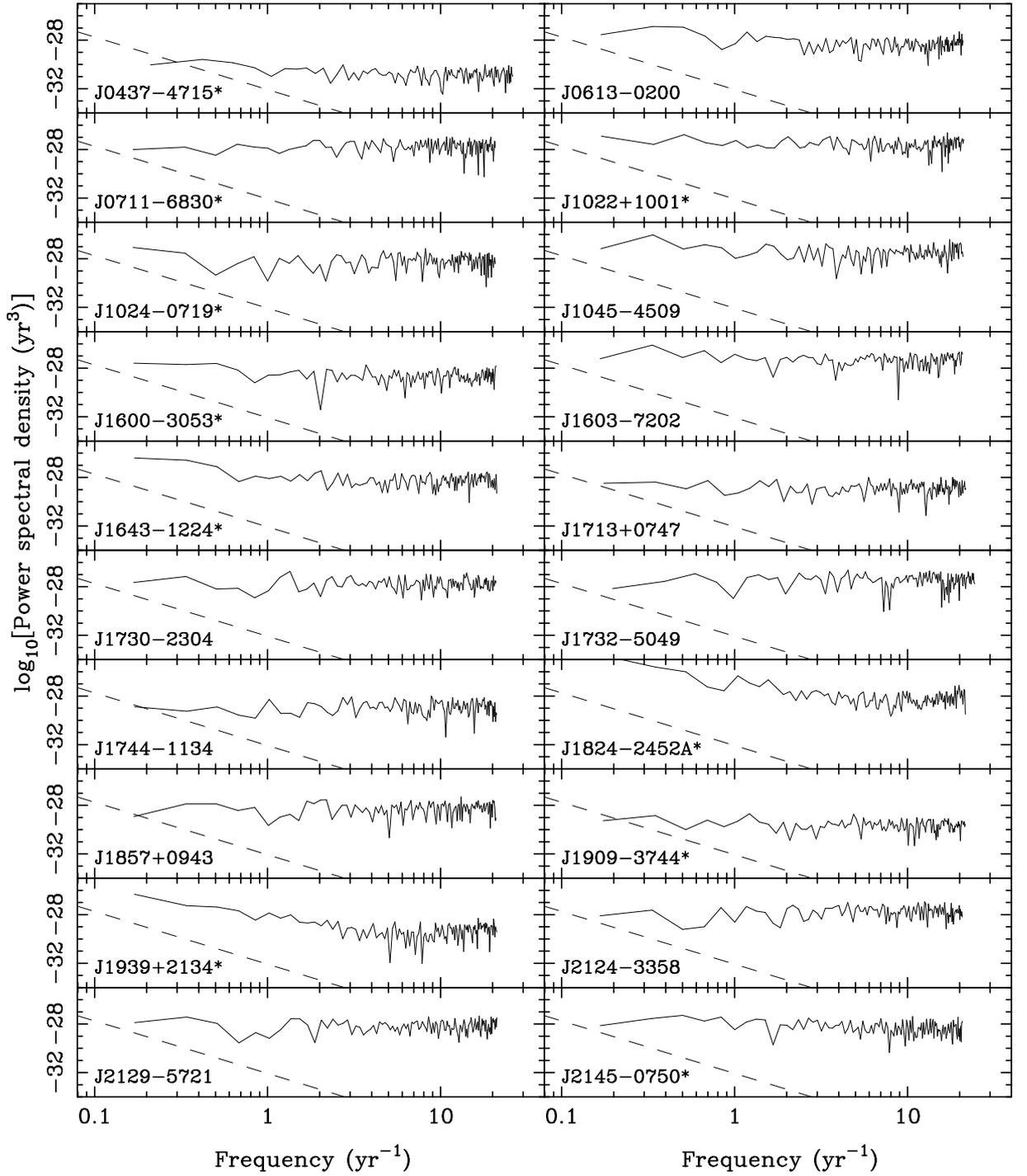}}
\caption{Power spectra of fluctuations in the timing residuals for the
  PPTA pulsars. The dashed line is the expected power spectrum in the
  timing residuals for a stochastic gravitational-wave background
  signal of amplitude $A_{\rm g} = 10^{-15}$. Note that the y-axis
  range is the same for all pulsars and covers nine orders of
  magnitude in power. Pulsars that have a $\ddot\nu$ with significance
  $\ge 3\sigma$ (Table~\ref{tb:f2}) are marked with *
  after the name. }\label{fg:spectra}
\end{figure*}

A more informative way to illustrate the pulse phase fluctuations for
a given pulsar is to compute the power spectrum of the timing
residuals. Figure~\ref{fg:spectra} shows such spectra for the PPTA
pulsars. These spectra were computed using a weighted least-squares
fit to the timing residuals shown in Figure~\ref{fg:dr1_res} after
whitening using the Cholesky method \citep{chc+11}. The line shown on
each of the spectral plots has a slope of $-13/3$ for $A_{\rm g} =
10^{-15}$ (Equation~\ref{eq:gwb_res}), a representative value for the
expected GW background from binary supermassive black holes in the
cores of distant galaxies \citep[e.g.,][]{svc08}. These plots
illustrate the wide differences in both red-noise and white-noise
properties between our pulsars. Consistent with the residual plots of
Figure~\ref{fg:dr1_res} and the rms residuals for the white noise
given in Table~\ref{tb:best}, the white-noise power level differs by
about four orders of magnitude from the ``best'' to the ``worst''
pulsars. More importantly, some pulsars have a much stronger red-noise
component than others. PSR J1824$-$2452A stands out with the highest
observed low-frequency noise level, most probably as a result of the
combined effects of intrinsic period irregularities and the varying
spatial accelerations due to gravitational interactions with other
stars in the globular cluster. PSR J1939+2134 also stands out, but
mainly because of its very precise ToAs; the absolute level of the red
noise in this pulsar is comparable to that of several others, e.g.,
PSRs J1045$-$4509, J1603$-$7202 and J1643$-$1224. The ten MSPs which
have $\ddot\nu$ significant at $3\sigma$ or greater in the best-band
fit (Table~\ref{tb:f2}) are marked with a * after the name in
Figure~\ref{fg:spectra}. Generally in these cases the power level in
the lowest-frequency bins is above the white-noise level.

It is important to note that, for the best pulsars, e.g., PSRs
J0437$-$4715, J1713+0747, J1857+0943 and J1909$-$3744, the observed
power level at the lowest frequency ($\sim 0.17$~yr$^{-1}$) is already
less than the power expected from a GW background with $A_{\rm g} =
10^{-15}$. However, there are some important caveats here. The GW
background line is the average for an infinite number of
universes. Any given realisation (e.g., our own Universe) may have a
higher or lower spectrum. In setting a limit on the actual GW
background, it is standard practice \citep[e.g.,][]{jhv+06} to take a
value for which 95\% of the realisations have a detection statistic
that is greater than the statistic for the observed spectrum. Also,
comparison of the GW prediction with the measured spectra assumes that
the Cholesky method compensates for the effects of fitting for $\nu$
and $\dot\nu$ on the post-fit residuals and their spectrum. The
accuracy of this compensation depends on the accuracy of the
estimation of the red-noise spectral model and this is difficult when
the red and white noise are of comparable amplitude. Despite these
caveats, we can safely conclude that the low-frequency noise level
seen in our best pulsars is already close to the average level
expected for an $A_{\rm g} = 10^{-15}$ stochastic background.

Conversely, it is clear that the red noise in the ``worst'' of our
pulsars cannot result from the GW background. If these signals were
from a GW background, they should be present in the timing residuals
of all pulsars and they are not. In fact, these red signals cannot
result from any ``global'' process such as timescale or solar-system
ephemeris errors which affect all pulsars. This is true even if there
are geometric factors that depend on the pulsar and/or source
position, since it is very unlikely that our best pulsars would all
happen to be located near the nulls of the geometric pattern. The most
likely sources of pulsar-specific red noise are intrinsic timing
irregularities and uncorrected interstellar variations.

These results lead to a key question for PTA projects: to what extent
will achievement of PTA goals be affected by intrinsic low-frequency
pulsar period noise as data spans increase?  Intrinsic fluctuations
are of course uncorrelated between different pulsars and so, with a
big enough pulsar sample, they will not preclude the reaching of PTA
objectives which inherently depend on the detection of correlated
signals. However there is no doubt that they make the task much more
difficult, especially if the intrinsic noise is of greater amplitude
than the target signal.

The degree to which extension of data spans helps GW detection efforts
depends critically on the spectral slope of the intrinsic red
noise. As Table~\ref{tb:f2} and Figure~\ref{fg:spectra} show, about
half of the PPTA pulsars have detectable red timing noise. Estimation
of the spectral properties of the red noise is difficult as, with a
couple of notable exceptions, viz, PSRs J1824$-$2452A and J1939+2134,
the red signal is only marginally above the white noise level even at
the lowest observed frequencies. Never-the-less, in order to make a
first estimate of the spectral properties, we fitted a power law plus
a constant (white) spectral model to the spectra shown in
Figure~\ref{fg:spectra}. Only four pulsars (J1024$-$0719,
J1643$-$1224, J1824$-$2452A and J1939+2134) had power-law spectral
indices greater than 2.5 and for only one of these (J1024$-$0719) was
the spectral index comparable to that of the expected GW spectrum. We
emphasise that these estimates have considerable uncertainty, but it
is clear that, for the majority of the PPTA MSPs, any red timing noise
detected so far has a spectrum that is much flatter than the expected
GW spectrum. We cannot preclude the presence of a currently invisible
red (non-GW) signal with a steep spectrum. However, with that proviso
and given current models for the origin of the GW background, we can
assert that within the next decade a GW signal should be dominating
the residuals of several of our pulsars and hence that a signficant
detection of that GW signal can be expected.

\subsection{The Extended PPTA Data Sets}\label{sec:dr1e}
The most important of the PTA objectives described in
\S\ref{sec:intro} are for detection of phenomena that generate
low-frequency fluctuations in residual time series, i.e., the spectra
of the expected signals are very red. Consequently, it is highly
advantageous to utilise data sets with long spans when searching for these
phenomena; this is discussed more quantitatively in \S\ref{sec:future}
below for the case of GW detection. The PPTA data sets for the 20
observed pulsars have data spans of approximately six years. However
Parkes timing data were obtained for most of these pulsars for up to
11 years prior to the start of the PPTA project
\citep{vbv+08,vbc+09}. In Appendix A, we present a re-analysis of
these data sets that enables them to be smoothly joined to the best
single-band PPTA data sets, giving data spans of up to 17.1 years.

\section{Future Prospects}\label{sec:future}
Figure~\ref{fg:spectra} shows that the spectral index for the expected
stochastic GW background is generally steeper than the spectral index
of other red processes that currently affect pulsar timing
residuals. Furthermore, most of these red processes are different for
different pulsars, whereas the variations due to the GW background
(and time-standard and ephemeris errors) are correlated. We can
therefore be confident that, given sufficiently long data spans and a
large enough sample of pulsars, PTA observations will reveal a GW
background unless current predictions of its strength at nanohertz
frequencies are incorrect.
 
\begin{figure*}
\centerline{\psfig{file=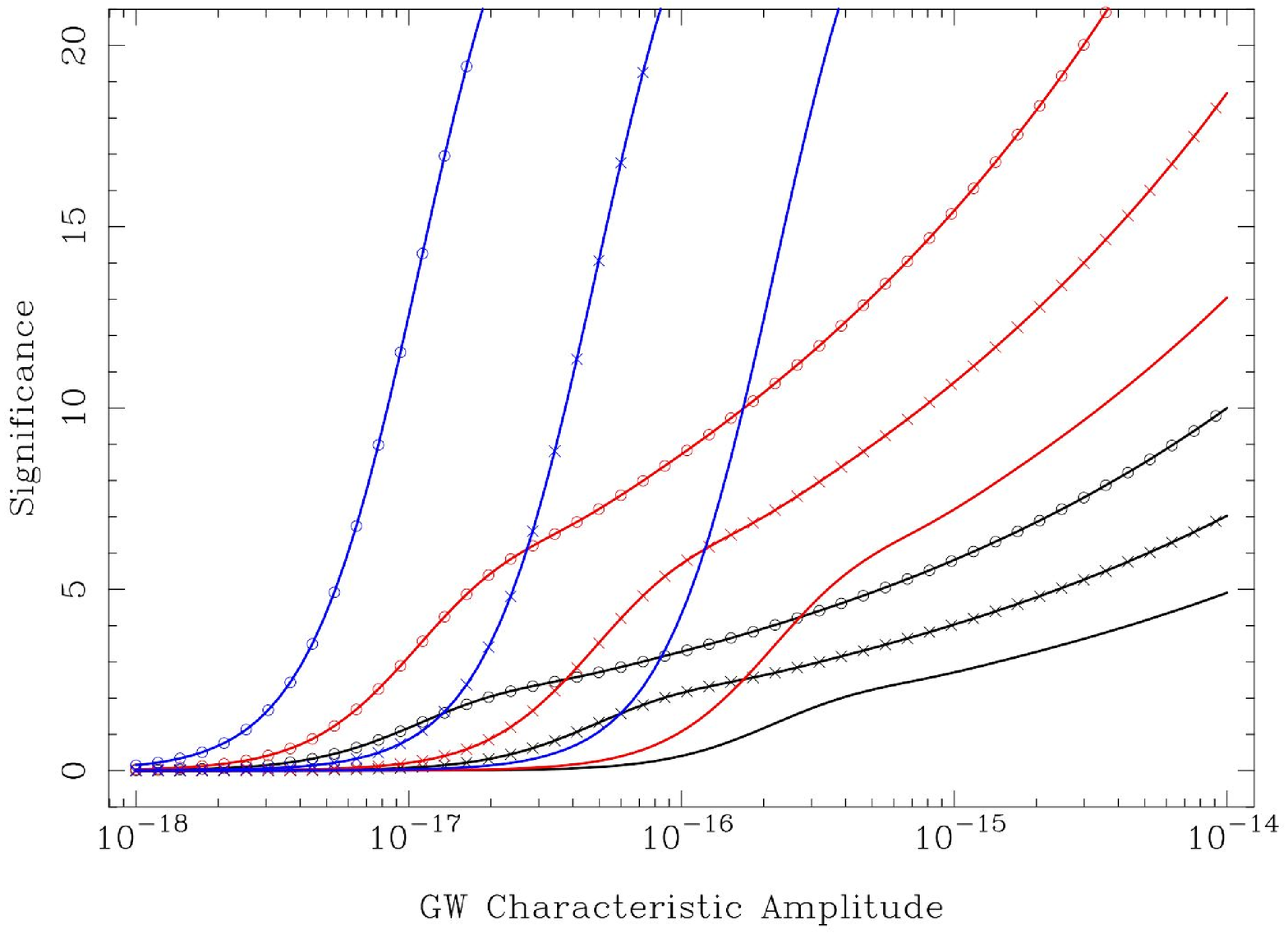,width=\textwidth}}
\caption{Sensitivity of a PTA to a stochastic background of GWs as a
  function of total data span $T_{\rm obs}$, number of pulsars $N$ in
  the PTA and assumming 100\,ns rms timing residuals. Black lines are
  for $N_{\rm psr}=20$ and, $T_{\rm obs}$ of 5 yr (unmarked), 10 yr
  ($\times$) and 20 yr ($\circ$), respectively, red lines are similar
  for $N_{\rm psr}=50$ and blue lines for $N_{\rm
    psr}=200$. }\label{fg:pta_sens}
\end{figure*}

Figure~\ref{fg:pta_sens} shows the dependence of PTA sensitivity on
the amplitude $A_{\rm g}$ (Equation~\ref{eq:gwb}) of a stochastic GW
background as a function of total data span $T_{\rm obs}$ and number
of pulsars in the PTA, $N_{\rm psr}$.  These curves were obtained
using Equation A12 of \citet{vbc+09}\footnote{Note that in
  \citet{vbc+09} Equation A8 the denominator should be squared and the
  number of spectral degrees of freedom should be $N_{\rm d.o.f.}
  \approx 1.4 T_{\rm obs}f_{\rm c}$ where $T_{\rm obs}$ is the data
  span and $f_{\rm c}$ is the corner frequency where the white-noise
  power and gravitational-wave power are equal. The reasons for this
  change are: a) at least at higher signal levels, quadratic fitting
  does not remove a degree of freedom from the fit, and b) the
  relevant bandwidth is that of the squared spectrum after Wiener
  filtering, that is, approximately $0.7 f_{\rm c}$.}  simulating an
ideal PTA with $N_{\rm psr}$ pulsars distributed randomly on the
sky, assuming weekly ToA sampling, 100~ns rms uncorrelated timing
residuals, and no ``red'' noise contributions from sources other than
the GW signal.

The lowest black line is the PPTA ``reference'' case discussed by
\citet{jhlm05}. Doubling the observing time decreases the GW amplitude
required for detection at a given significance level by a factor of
approximately $4.5\sim 2^{13/6}$ (cf., Equation~\ref{eq:gwb_res},
remembering that the significance is in terms of amplitude rather than
power). The effect of doubling observing time on significance at a
given amplitude greatly depends on the level of the GW amplitude;
generally it is smaller for larger amplitudes because the GWs passing
over the pulsars contribute an uncorrelated noise term, thereby
reducing the significance. Increasing the number of pulsars in the PTA
improves the significance at a given GW amplitude by approximately
$[N_{\rm psr}(N_{\rm psr}-1)]^{1/2} \sim N_{\rm psr}$. Increasing the
white noise rms amplitude by a factor of five, to 500~ns, has almost
exactly the same effect as halving the data span, i.e., 10 years with
rms residual of 500~ns gives essentially the same sensitivity curve as
5 years with 100~ns rms residual. Given the sensitivity we can achieve
with current instruments, the 10-year -- 500-ns scenario is more
realistic for the PPTA to achieve the necessary detection sensitivity. 

Improving rms timing residuals will also help, but this is difficult
with present telescope facilities. In the Northern Hemisphere there
are a number of radio telescopes with collecting area larger than that
of the Parkes telescope which should give better results given similar
instrumentation and similar observing cadence. For Parkes, a more
sensitive broad-band receiver is currently under development and this
will give improved results.

However, as Figure~\ref{fg:pta_sens} makes clear, the greatest benefit
is obtained by increasing the number of pulsars in the PTA. On-going
searches \citep[e.g.,][]{kjb+12} have already found MSPs which are
suitable for PTAs and the best of these have already been included in
existing PTAs. In time these will contribute fully to PTA
sensitivity. This approximately linear increase in PTA sensitivity with
$N_{\rm psr}$ is the strongest motivation to combine the results from
the three existing (and possible future) PTA projects. The EPTA,
NANOGrav and the PPTA are collaborating to form the International
Pulsar Timing Array (IPTA) \citep{haa+10}. Although there is
considerable overlap between the samples of the three main existing
PTA projects, \citet{haa+10} list 37 pulsars that are regularly
observed by at least one PTA. Of course, the data quality varies
considerably across this sample, but there is no doubt that an IPTA
data set will give superior results compared to the data set of any
one participating PTA. Even at the pessimistic end of current
predictions \citep[e.g.,][]{svc08}, IPTA data sets with spans of 10
years or more should give a significant detection of the GW background
from binary black holes in distant galaxies. Failure to detect the GW
background with these data sets will imply that current models for
this background are fundamentally flawed. This would have important
implications for current theories of black-hole formation and
evolution in galaxies and for theories of galaxy growth through
mergers.

Even if a detection of GWs is achieved with IPTA data sets, it is
clear that detailed study of the properties of nanohertz GWs and their
sources will have to wait until the Square Kilometer Array (SKA)
\citep[see, e.g.,][]{ckl+04} is operational. The enormous increase in
sensitivity provided by the SKA will allow detection and subsequent
frequent timing observations of a large sample of MSPs. Ultimately a sample
of 200 PTA-quality MSPs should be possible, although it will
take time to build up a significant data span. Once that is done,
Figure~\ref{fg:pta_sens} shows that high-significance detetections
will be possible even if the GW background is relatively weak,
enabling accurate measurement of properties such as spectral index and
possible origin of the background \citep[e.g.,][]{vlj+11} and
investigation of non-GR effects in the received signals
\citep[e.g.,][]{ljp08}.

Although Figure~\ref{fg:pta_sens} is computed for the case of
detection of a stochastic GW background, similar considerations apply
to other PTA objectives. Improved data sets should enable the
detection and study of individual GW sources, for example, bright
black-hole binaries \citep{svv09} or burst signals from cosmic strings
\citep{dv05,sbs12}. Source locations with arc-minute precision or
better should be measurable and other source properties can be studied in
detail \citep{lwk+11}. Extended data sets give improved
pulsar-based timescales \citep{hcm+12}, enable improved determinations
of the masses of solar-system planets and possibly even allow the
detection of currently unknown solar-system objects such as TNOs. The
future prospects of PTA research are indeed exciting.

\section*{Acknowledgments}
The PPTA project was initiated with support from RNM's Australian
Research Council (ARC) Federation Fellowship (FF0348478) and from the
CSIRO under this Fellowship program. It has also received support from
ARC Discovery Project grant DP0985272. GH is the recipient of an ARC
QEII Fellowship (DP0878388) and VR is a recipient of a John Stocker
Postgraduate Scholarship from the Science and Industry Endowment
Fund. Part of this research was carried out at the Jet Propulsion
Laboratory, California Institute of Technology, under a contract with
the National Aeronautics and Space Administration. We acknowledge
contributions to the project from L. Kedziora-Chudczer, K. J. Lee,
A. N. Lommen, D. Smith and Ding Chen. The Parkes telescope is part of
the Australia Telescope which is funded by the Commonwealth Government
for operation as a National Facility managed by CSIRO.


\clearpage
\section*{Appendix A -- Extension of the PPTA Data Sets}
Many of the PPTA pulsars had been observed at Parkes for some years
prior to the commencement of the PPTA project
\citep[eg.,][]{tsb+99,hbo06}. Long data spans are very beneficial to
many pulsar timing studies and so it is useful to combine these
earlier data with the PPTA observations. Here we describe the
``Extended PPTA'' data sets that add earlier observations
as described by \citet{vbv+08,vbc+09} to the PPTA data sets.

\citet{vbv+08} analysed Parkes observations of PSR J0437$-$4715
obtained over a 10-year period from 1996 to 2006 using two different
receivers (H-OH and multibeam centre beam) operating in the 20cm band
and with four different back-end systems operating at different
times. With these data, they obtained a precise distance for the
pulsar, both with a direct measurement of the annual parallax and by
measuring the apparent orbital period derivative resulting from the
transverse motion of the system. The back-end systems used were the
Fast Pulsar Timing Machine (FPTM) \citep{sbm+97} in 1996 -- 1997, the
S2 VLBI recorder \citep{van03a} in 1997 -- 1998, the
Caltech-Parkes-Swinburne recorder CPSR \citep{van03a} for 1998 --
2002 and CPSR2 for 2002 -- 2006.

Similarly, \citet{vbc+09} presented 20cm data for all 20 PPTA pulsars
with data spans ranging from 3.9 years to 14.2 years recorded between
1994 and 2008. The FPTM was used from 1992 to 2001, a 512-channel
analogue filterbank \citep{dlm+01} in 2002 and 2003 and CPSR2 from
2002 November to 2008. For three pulsars, PSRs J1045$-$4509,
J1909$-$3744 and J1939+2134, data from the 50cm receiver was used to
model the DM variations after this receiver became available in 2003
November.

These data were analysed by \citet{vbv+08,vbc+09} using {\sc Tempo2}
and fitting for the pulsar astrometric, pulse frequency and binary
parameters (if appropriate) simultaneously with arbitrary offsets
between the different instruments. As \citet{ych+11} have shown,
fitting for offsets (or jumps) subtracted most of the low-frequency
power from the post-fit residuals. We have therefore reanalysed the
\citet{vbv+08,vbc+09} data sets to separately determine and fix as
many of the offsets as possible. Where overlapping or near-overlapping
(gap $\lapp 150$~d) data from two different instruments were
available, the offsets were measured from short data spans (typically
a few months) fitting for just $\nu$ and the offset with all other
parameters held at the values from a fit to the entire data span for
that pulsar. Offsets determined in this way were held fixed in
subsequent analyses. For a few pulsars, data gaps between some pairs
of instruments were too large for this procedure to give reliable
results and arbitrary jumps were retained.

The \citet{vbv+08,vbc+09} data sets were then combined with the
three-band PPTA data sets described in the main text to form extended
PPTA data sets with data spans of up to 17.1 years. These extended
data sets were then analysed using {\sc Tempo2} with the Cholesky
decomposition \citep{chc+11} to properly handle the red noise
component. The DM and DM offsets determined as described in
\S\ref{sec:dmv} for the PPTA data spans were included and held
fixed. Since most of the \citet{vbv+08,vbc+09} observations were at
frequencies in or close to the 20cm band, it was generally impossible
to obtain DM variations from these data that were sufficiently precise
to enable correction for variable dispersion delays. The exception is
PSR J1939+2134, where dual-band FPTM observations (band centres about
1420 MHz and 1650 MHz) enabled a DM offset at MJD 50350 (1996
September) to be measured. This was held fixed in the subsequent
Cholesky fit along with the other DM offsets.

For all pulsars, parameters for which significant values could be
obtained were included in the fit. Finally, the remaining instrumental
offsets were included. The use of the Cholesky method ensures that
these offsets (along with the other parameters) and their
uncertainties are reliably determined despite the presence of the red
noise. The first seven columns of Table~\ref{tb:dr1e} give the data
spans in MJD and years, the total number of ToAs (all bands), number
of pulsar parameters fitted (N$_{\rm p}$), the number of fixed
instrumental offsets or jumps (N$_{\rm j0}$) and the number of
instrumental offsets included in the final fit (N$_{\rm j1}$). The
fixed instrumental jumps include three interband jumps; the ``best''
band is designated as reference and the other three bands (including
both 40cm and 50cm) are referred to it. Fitted jumps were needed for
only seven pulsars and for all of these except PSR J0437$-$4715, the
number required was three or less. PSR J0437$-$4715 required more
offsets because more instruments were used for observations of this
pulsar and the high precision of the ToAs revealed a larger number of
significant offsets.

\begin{table*}[h]
\begin{center}
\caption{The Extended PPTA data sets}\label{tb:dr1e}
\begin{tabular}{lcd{3}d{2}d{1}d{0}d{0}d{0}d{5}cc}
\hline 
\multicolumn{1}{c}{PSR} & \multicolumn{2}{c}{Data Span} & \multicolumn{1}{c}{N$_{\rm ToA,3B}$} & \multicolumn{1}{c}{N$_{\rm p}$} & \multicolumn{1}{c}{N$_{\rm j0}$} & \multicolumn{1}{c}{N$_{\rm j1}$} & \multicolumn{1}{c}{N$_{\rm ToA}$} & \multicolumn{1}{c}{Rms Res.} &$\chi^2_{\rm r}$ &  $\ddot\nu$ \\
&  (MJD) & \multicolumn{1}{c}{(yr)} & & & & & & \multicolumn{1}{c}{($\mu$s)} & & ($10^{-28}$ s$^{-3}$) \\
\hline 
J0437$-$4715 & 50190 -- 55619& 14.86 &5055&15 &7 &9 & 3508 & 0.21 & 7.17 & $1.26\pm0.03$\\
J0613$-$0200 & 51526 -- 55619& 11.21 &629 &12 &6 &0 & 341 & 1.11 & 1.24 & $7.2\pm2.1$  \\
J0711$-$6830 & 49373 -- 55620& 17.10 &555 &6  &10&1 & 319 & 1.54 & 1.54 & $-0.8\pm0.7$ \\
J1022+1001   & 52649 -- 55619&  8.13 &624 &12 &7 &0 & 378 & 1.82 & 8.14 & $-0.4\pm1.6$ \\
J1024$-$0719 & 50117 -- 55620& 15.07 &493 &6  &10&0 & 309 & 4.38 & 12.57 & $-38.6\pm0.8$ \\
J1045$-$4509 & 49405 -- 55620& 17.02 &635 &10 &10&0 & 393 & 5.05 & 3.18 & $9.3\pm1.2$ \\
J1600$-$3053 & 52301 -- 55598&  9.03 &704 &12 &7 &1 & 503 & 0.98 & 1.21 & $8.6\pm2.2$ \\
J1603$-$7202 & 50026 -- 55619& 15.31 &483 &12 &7 &3 & 290 & 2.12 & 3.08 & $1.2\pm0.4$ \\
J1643$-$1224 & 49421 -- 55598& 16.91 &477 &11 &7 &3 & 288 & 2.30 & 5.90 & $6.0\pm1.0$ \\
J1713+0747   & 49421 -- 55619& 16.97 &612 &15 &10&0 & 334 & 0.46 & 7.75 & $-2.60\pm0.16$ \\
J1730$-$2304 & 49421 -- 55598& 16.91 &390 &7  &10&0 & 223 & 2.59 & 3.25 & $-0.8\pm0.8$ \\
J1732$-$5049 & 52647 -- 55582&  8.04 &244 &11 &9 &0 & 149 & 2.47 & 1.17 & $28\pm7$ \\
J1744$-$1134 & 49729 -- 55599& 16.07 &534 &7  &9 &3 & 368 & 0.65 & 3.27 & $1.9\pm0.3$ \\
J1824$-$2452A& 53518 -- 55620&  5.75 &302 &6  &3 &0 & 178 & 2.02 & 14.50 & $241\pm22$ \\
J1857+0943   & 53086 -- 55599&  6.88 &291 &15 &7 &0 & 152 & 0.96 & 1.18 & $7.1\pm5.2$ \\
J1909$-$3744 & 52618 -- 55619&  8.22 &1245&14 &7 &0 & 724 & 0.19 & 5.06 & $3.54\pm0.44$ \\
J1939+2134   & 49956 -- 55599& 15.45 &386 &7  &9 &3 & 237 & 4.27 & 3664 & $127.8\pm1.4$ \\
J2124$-$3358 & 49489 -- 55619& 16.78 &652 &7  &11&0 & 473 & 2.92 & 1.85 & $-6.1\pm0.8$ \\
J2129$-$5721 & 49987 -- 55619& 15.42 &448 &11 &10&0 & 285 & 1.41 & 2.21 & $6.3\pm0.8$ \\
J2145$-$0750 & 49517 -- 55618& 16.70 &972 &13 &10&0 & 696 & 1.06 & 2.81 & $-1.38\pm0.12$ \\
\hline
\end{tabular}
\end{center}
\end{table*}

\begin{figure*}[h]
\centerline{\psfig{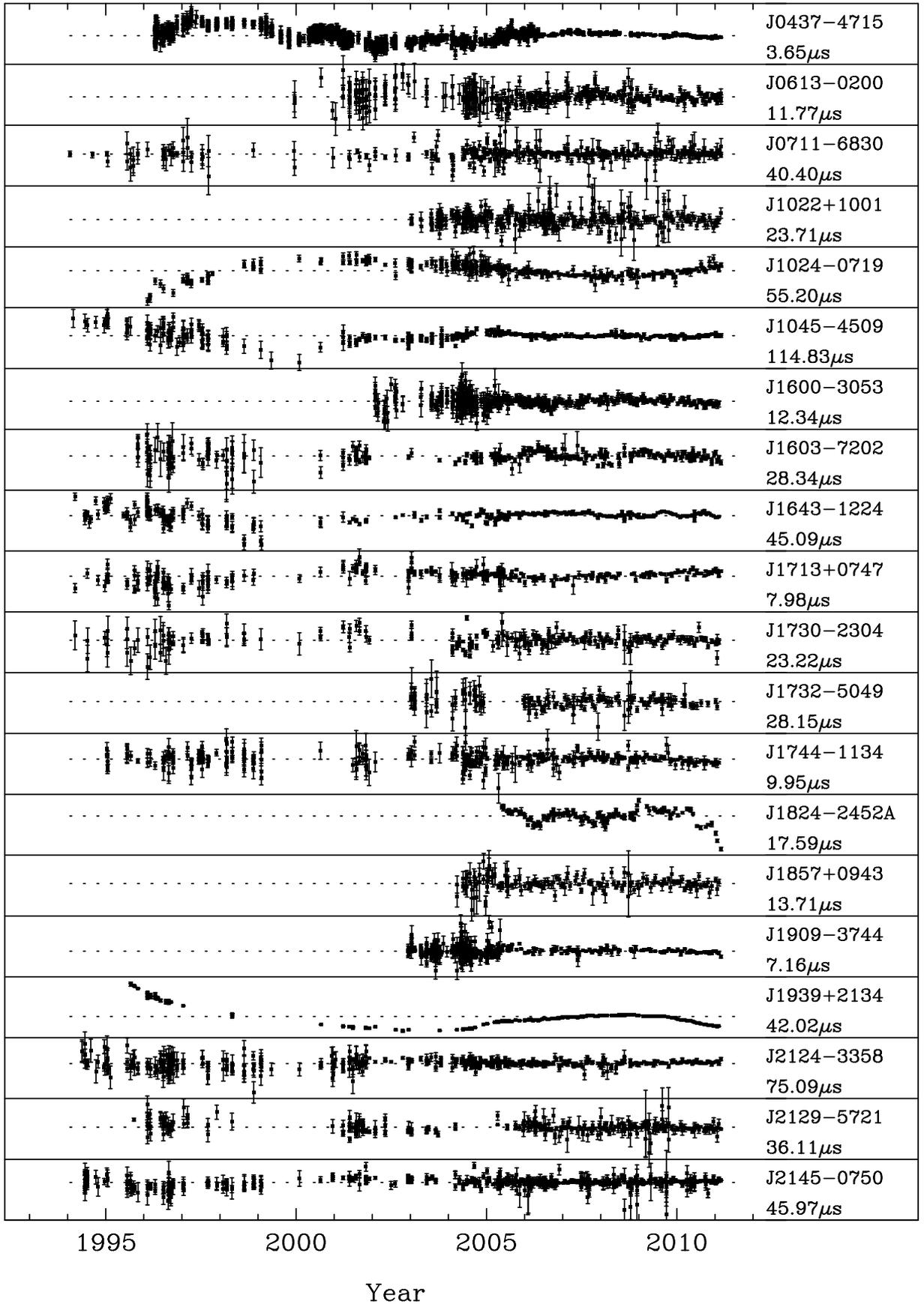}}
\caption{Post-fit timing residuals for the extended PPTA data
  sets. The vertical extent of each subplot is adjusted to fit the
  data and its value is given below the pulsar name. The dashed line
  marks zero residual.}\label{fg:dr1e_res}
\end{figure*}

To form a ``best'' extended data set, the \citet{vbv+08,vbc+09} data
sets were combined with the best PPTA data set described in
\S~\ref{sec:best} and Table~\ref{tb:best}. Parameter files were
created by copying the pulsar parameters and jumps and DM corrections
for the extended three-band fit described above and holding all
parameters except $\nu$ and $\dot\nu$ fixed.  Columns 8 -- 10 of
Table~\ref{tb:dr1e} give the number of ToAs in the best single-band
extended data set, and the post-fit rms timing residual and reduced
$\chi^2$ for each pulsar. Note that Cholesky decomposition was not
used in this fit. Figure~\ref{fg:dr1e_res} shows the post-fit
residuals, clearly illustrating the better quality of the PPTA
data sets compared to those obtained with the earlier systems. For
several pulsars, e.g., PSRs J0437$-$4715 and J1024$-$0719, they also
reveal red noise that was obvious in neither the 6-year PPTA data sets
(Figure~\ref{fg:dr1_res}), nor in the post-fit residuals presented by
\citet{vbv+08,vbc+09}. 

To quantify these in the same way as for the PPTA data sets
(Table~\ref{tb:f2}), $\ddot\nu$ was additionally fitted, giving the
values in the last column of Table~\ref{tb:dr1e}. In contrast to the
PPTA data sets where ten of the 20 pulsars had $\ddot\nu$ values
significant at the $3\sigma$ level, all but three of the pulsars have
significant $\ddot\nu$ values for the extended data set. This is a
simple illustration of the importance of long data spans for
characterising red-noise processes. PSR J1024$-$0719 is a particularly
striking example: for the PPTA data set this had a $\ddot\nu$ with a
significance of $5\sigma$, whereas for the extended data set the value
is consistent but the significance has increased to $48\sigma$. It is
interesting to note that, based on the PPTA data set (\S\ref{sec:f2}),
this pulsar appears to have a relatively steep red timing noise
spectrum.

PSRs J1824$-$2452A and J1939+2134 stand out with positive and
relatively large $\ddot\nu$ values, but it is notable that 13 of the
17 significant $\ddot\nu$ in Table~\ref{tb:f2} are also positive. A
small glitch was observed for PSRs J1824$-$2452A in 2001
\citep{cb04}. It is well known that large positive values of
$\ddot\nu$ are observed after glitches in young pulsars
\citep{lsg00,ywml10}. These are attributed to the dynamics of
superfluid vortices in the interior of the neutron star as they
re-establish equilibrium with the rigid crust following a glitch
\citep[e.g.,][]{accp93}.  This suggests the intriguing possibility
that the observed positive $\ddot\nu$ values are attributable to
post-glitch recoveries, in all cases, except possibly for PSR
J1824$-$2452A, from unseen earlier glitches. It is worth noting that
the $\ddot\nu$ values seen for the PPTA MSPs are about four orders of
magnitude smaller than those seen in young pulsars.

Comparison of the observed rms timing residuals for the extended data
sets with the ``CP+MSP'' model of \citet{sc10} 
gives similar results to those for the PPTA data sets discussed in
\S\ref{sec:f2}. Compared to the PPTA data sets, the $\zeta$ values for
most pulsars are a factor of 2 -- 8 smaller, indicating the diminished
relative contribution of the white noise with the longer data spans. For PSR
J0437$-$4715, $\zeta\sim 0.15$ showing that the model substantially
over-predicts the amount of timing noise with this longer data
span. For PSRs J1909$-$3744 and J1939+2134, $\zeta$ values are 1.18
and 0.49 respectively, similar to those for the shorter data spans; for
PSR J1824$-$2452A there are no new data in the extended data set. 

For several pulsars, most notably PSRs J0437$-$4715 and J1713+0713,
there is significant red noise in the \citet{vbv+08,vbc+09} data sets
alone which was not present in the results presented by
\citet{vbv+08,vbc+09}. This is mostly attributable to the fixing of
the instrumental offsets. However, it is likely that the inability to
accurately model DM variations for most of the early data also
contributes significantly to the observed residuals. For example, the
deviations observed in the early data for PSR J1045$-$4509 are almost
certainly due to uncorrected DM variations \citep[cf.,][]{kcs+12}. It
is possible that observations from other observatories will help to
determine pre-PPTA DM variations. However it is generally difficult if
not impossible to establish absolute ToAs for archival data with
sufficient precision and so these data are likely to be of limited use
in this respect. Only future carefully calibrated multi-band
observations will establish for sure if DM variations are a
significant contributor to the observed red noise on decade-long
timescales.

\end{document}